\documentclass[review]{elsarticle}

\usepackage{hyperref}

\usepackage{amsmath}
\usepackage{array}
\usepackage[english]{babel}
\usepackage[justification=justified]{caption}
\usepackage{graphicx}
\usepackage[utf8]{inputenc}
\usepackage{makecell,multirow,longtable, lscape}
\usepackage{pifont}
\usepackage{rotating}
\usepackage{tabularx}
\usepackage{todonotes}
\usepackage{xcolor}

\newcolumntype{M}[1]{>{\centering\arraybackslash} m{#1}}
\newcolumntype{Y}{>{\centering\arraybackslash}X}

\definecolor{myBlue}{HTML}{4682B4}

\newcommand{\xmark}{\text{\ding{55}}}
\usepackage{amssymb}
\usepackage{xcolor}
\newif\ifshowcomments
\showcommentstrue

\ifshowcomments
\newcommand{\mynote}[2]{\fbox{\bfseries\sffamily\scriptsize{#1}}
	{\small$\blacktriangleright$\textsf{\emph{#2}}$\blacktriangleleft$}}
\else
\newcommand{\mynote}[2]{}
\fi

\journal{Computers and Security}

\begin{document}

\begin{frontmatter}

\title{Survey of Machine Learning Techniques\\for Malware Analysis}
\author[addr1]{Daniele Ucci\corref{mycorrespondingauthor}}
\ead{ucci@diag.uniroma1.it}
\author[addr2]{Leonardo Aniello}
\ead{l.aniello@soton.ac.uk}
\author[addr1]{Roberto Baldoni}
\ead{baldoni@diag.uniroma1.it}

\address[addr1]{Research Center of Cyber Intelligence and Information Security, ``La Sapienza'' University of Rome}
\address[addr2]{Cyber Security Research Group, University of Southampton}

\begin{abstract}

Coping with malware is getting more and more challenging, given their relentless growth in complexity and volume.
One of the most common approaches in literature is using machine learning techniques, to automatically learn models and patterns behind such complexity, and to develop technologies to keep pace with malware evolution.
This survey aims at providing an overview on the way machine learning has been used so far in the context of malware analysis in Windows environments, i.e. for the analysis of Portable Executables.
We systematize surveyed papers according to their objectives (i.e., the \textit{expected output}), what information about malware they specifically use (i.e., the \textit{features}), and what machine learning techniques they employ (i.e., what \textit{algorithm} is used to process the input and produce the output).
We also outline a number of issues and challenges, including those concerning the used \textit{datasets}, and identify the main current topical trends and how to possibly advance them. In particular, we introduce the novel concept of \textit{malware analysis economics}, regarding the study of existing trade-offs among key metrics, such as analysis accuracy and economical costs.
\end{abstract}

\begin{keyword}
portable executable, malware analysis, machine learning, benchmark, malware analysis economics
\end{keyword}

\end{frontmatter}

\section{Introduction} \label{sec:intro}

Despite the significant improvement of cyber security mechanisms and their continuous evolution, malware are still among the most effective threats in the cyber space. %
Malware analysis applies techniques from several different fields, such as program analysis and network analysis, for 
the study of malicious samples to develop a deeper understanding on several aspects, including their behaviour and how they evolve over time. %
Within the unceasing arms race between malware developers and analysts, each advance in security technology is usually promptly followed by a corresponding evasion. 
Part of the effectiveness of novel defensive measures depends on what properties they leverage on. %
For example, a detection rule based on the MD5 hash of a known malware can be easily eluded by applying standard techniques like obfuscation, or more advanced approaches such as \textit{polymorphism} or \textit{metamorphism}. For a comprehensive review of these techniques, refer to Ye \textit{et al.}~\cite{Ye2017}. These methods change the binary of the malware, and thus its hash, but leave its behaviour unmodified. 
On the other side, developing detection rules that capture the semantics of a malicious sample is much more difficult to circumvent, because malware developers should apply more complex modifications.
A major goal of malware analysis is to capture additional properties to be used to improve security measures and make evasion as hard as possible.
Machine learning is a natural choice to support such a process of knowledge extraction.
Indeed, many works in literature have taken this direction, with a variety of approaches, objectives and results.

\bigskip

This survey aims at reviewing and systematising existing literature where machine learning is used to support malware analysis of Windows executables, i.e. Portable Executables (PEs).
The intended audience of this survey includes any security analysts, i.e. security-minded reverse engineer or software developer, who may benefit from applying machine learning to automate part of malware analysis operations and make the workload more tractable.
Although mobile malware represents an ever growing threat, Windows largely remains the preferred target~\cite{av-test-report-2017} among all the existing platforms. Malware analysis techniques for PEs are slightly different from those for Android apps because there are significant dissimilarities on how operating system and applications work. %
As a matter of fact, literature papers on malware analysis commonly point out what specific platform they target, so we specifically focus on works that consider the analysis of PEs.
64 recent papers have been selected on the basis of their bibliographic significance, reviewed and systematised according to a taxonomy with three fundamental dimensions: (i) the specific \textit{objective} of the analysis, (ii) what types of \textit{features} extracted from PEs they consider and (iii) what machine learning \textit{algorithm}s they use.
We distinguish three main objectives: \textit{malware detection}, \textit{malware similarity analysis} and %
\textit{malware category detection}. %
PE features have been grouped in eight types: 
\textit{byte sequences}, 
\textit{APIs/System calls}, 
\textit{opcodes},  
\textit{network}, 
\textit{file system}, 
\textit{CPU registers}, 
\textit{PE file characteristics} and  \textit{strings}. %
Machine learning algorithms have been categorized depending on whether the learning is \textit{supervised}, \textit{unsupervised} or \textit{semi-supervised}.
The characterisation of surveyed papers according to such taxonomy allows to spot research directions that have not been investigated yet, such as the impact of particular combination of features on analysis accuracy.
The analysis of such a large literature leads to single out three main issues to address. The first concerns overcoming modern anti-analysis techniques such as encryption. The second regards the inaccuracy of malware behaviour modelling due to the choice of what operations of the sample are considered for the analysis. The third is about the obsolescence and unavailability of the datasets used in the evaluation, which affect the significance of obtained results and their reproducibility. In this respect, we propose a few guidelines to prepare suitable benchmarks for malware analysis through machine learning.
We also identify a number of topical trends that we consider worth to be investigated more in detail, such as malware attribution and triage. %
Furthermore, we introduce the novel concept of \textit{malware analysis economics}, regarding the existing trade-offs between analysis accuracy, time and cost, which should be taken into account when designing a malware analysis environment. %

The novel contributions of this work are
\begin{itemize}
	\item the definition of a taxonomy to synthesise the state of the art on machine learning for malware analysis of PEs;
	\item a detailed comparative analysis of existing literature on that topic, structured according to the proposed taxonomy, which highlights possible new research directions;
	\item the determination of present main issues and challenges on that subject, and the proposal of high-level directions to investigate to overcome them;
	\item the identification of a number of topical trends on machine learning for malware analysis of PEs, with general guidelines on how to advance them;
	\item the definition of the novel concept of malware analysis economics.
\end{itemize}

\bigskip

The rest of the paper is structured as follows. 
Related work are described in Section~\ref{sec:rel_work}.
Section~\ref{sec:taxonomy} presents the taxonomy we propose to organise reviewed malware analysis approaches based on machine learning, which are then characterised according to such a taxonomy in Section~\ref{sec:characterization}.
From this characterisation, current issues and challenges are pointed out in Section~\ref{sec:issues_challenges}.
Section~\ref{sec:topical_trends} highlights topical trends and how to advance them. Malware analysis economics is introduced in Section~\ref{sec:malware_analysis_economics}.
Finally, conclusions and future works are presented in Section~\ref{sec:conclusion}.

\section{Related Work} \label{sec:rel_work}

Other academic works have already addressed the problem of surveying contributions on the usage of machine learning techniques for malware analysis. 
The survey written by Shabtai \textit{et al.}~\cite{Shabtai:2009:DMC:1550969.1551289} is the first one on this topic. It specifically deals with how classifiers are used on static features to detect malware. 
As most of the other surveys mentioned in this subsection, the main difference with our work is that our scope is wider as we target other objectives besides malware detection, such as similarities analysis and category detection. Furthermore, a novel contribution we provide is the idea of malware economics, which is not mentioned by any related work.
Also in~\cite{SahuAhirwarHemlata2014}, the authors provide a comparative study on papers using pattern matching to detect malware, by reporting their advantages, disadvantages and problems. 
Souri and Hosseini~\cite{Souri2018} proposes a taxonomy of malware detection approaches based on machine learning. In addition to consider detection only, their work differs from ours because they do not investigate what features are taken into account. 
LeDoux and Lakhotia~\cite{LeDoux2015} describe how machine learning is used for malware analysis, whose end goal is defined there as ``automatically detect malware as soon as possible, remove it, and repair any damage it has done''. 

Bazrafshan \textit{et al.}~\cite{Bazrafshan2013} focus on malware detection and identify three main methods for detecting malicious software, i.e. based on signatures, behaviours and heuristics, the latter using also machine learning techniques. They also identify what classes of features are used by reviewed heuristics for malware detection, i.e. API calls, control flow graphs, n-grams, opcodes and hybrid features. In addition to going beyond malware detection, we propose a larger number of feature types, which reflects the wider breadth of our research.

Basu \textit{et al.}~\cite{Basu2016} examine different works relying on data mining and machine learning techniques for the detection of malware. They identify five types of features: API call graph, byte sequence, PE header and sections, opcode sequence frequency and kernel, i.e. system calls. In our survey we establish more feature types, such as strings, file system and CPU registers. They also compare surveyed papers by used features, used dataset and mining method.

Ye \textit{et al.}~\cite{Ye2017} examine different aspects of malware detection processes, focusing on feature extraction/selection and classification/clustering algorithms. Also in this case, our survey looks at a larger range of papers by also including many works on similarity analysis and category detection. They also highlight a number of issues, mainly dealing with machine learning aspects (i.e. incremental learning, active learning and adversarial learning). We instead look at current issues and limitations from a distinct angle, indeed coming to a different set of identified problems that complement theirs.  Furthermore, they outline several trends on malware development, while we rather report on trends about machine learning for malware analysis, again complementing their contributions.

Barriga and Yoo~\cite{Barriga2017MalwareDA} briefly survey literature on malware detection and malware evasion techniques, to discuss how machine learning can be used by malware to bypass current detection mechanisms.
Our survey focuses instead on how machine learning can support malware analysis, even when evasion techniques are used.
Gardiner and Nagaraja~\cite{Gardiner:2016:SML:2988524.3003816} concentrate their survey on the detection of command and control centres through machine learning.

\section{Taxonomy of Machine Learning Techniques for Malware Analysis} \label{sec:taxonomy}

This section introduces the taxonomy on how machine learning is used for malware analysis in the reviewed papers.
We identify three major dimensions along which surveyed works can be conveniently organised. The first one characterises the final \textit{objective} of the analysis, e.g. malware detection. The second dimension describes the \textit{features} that the analysis is based on in terms of how they are extracted, e.g. through dynamic analysis, and what features are considered, e.g. CPU registers. Finally, the third dimension defines what type of \textit{machine learning algorithm} is used for the analysis, e.g. supervised learning.

\begin{figure}
	\centering
	\includegraphics[width=\linewidth]{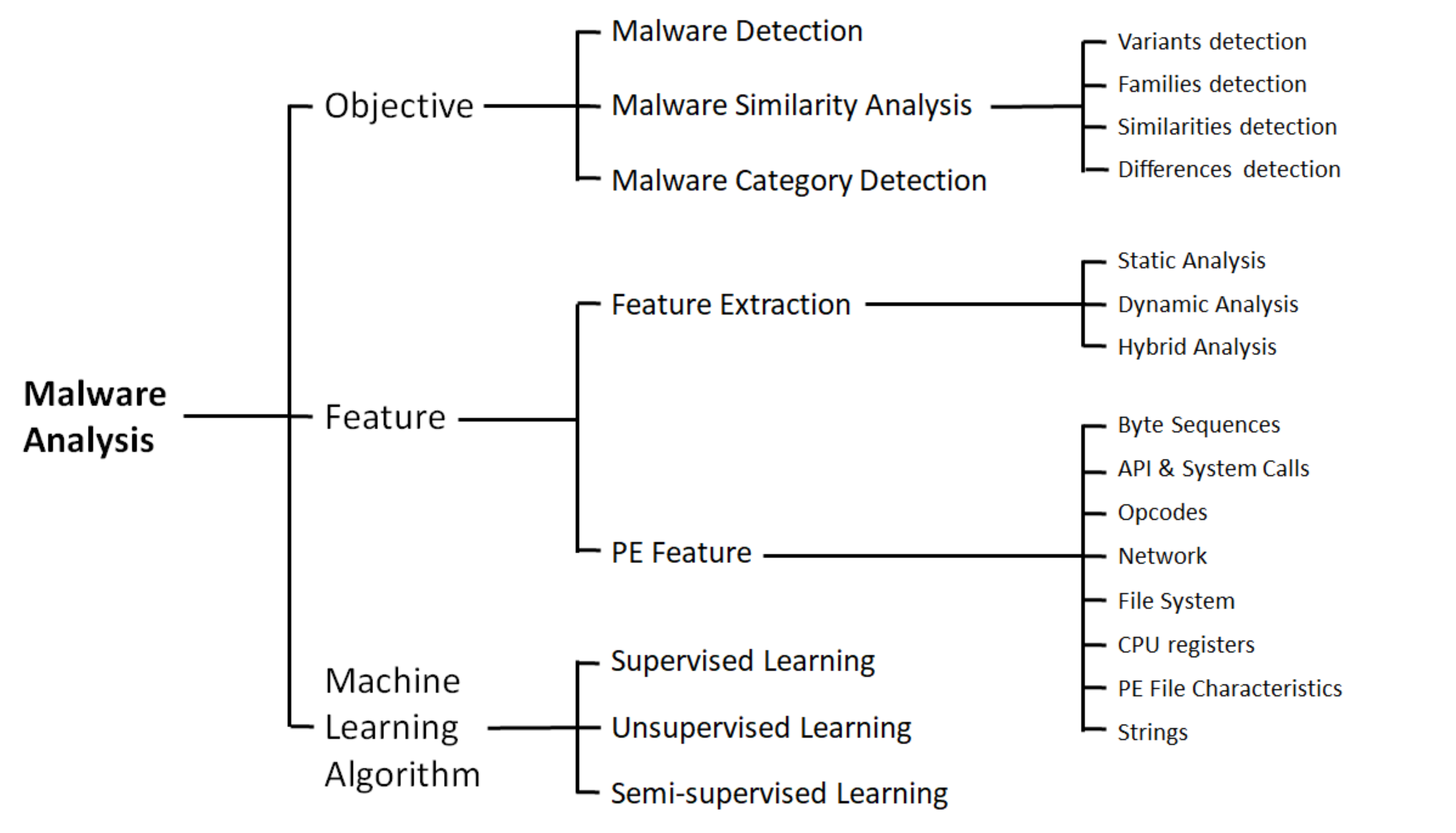}
	\captionof{figure}{Taxonomy of machine learning techniques for malware analysis}
	\label{fig:taxonomy}
\end{figure}

Figure~\ref{fig:taxonomy} shows a graphical representation of the taxonomy. The rest of this section is structured according to the taxonomy. Subsection~\ref{sec:taxonomy_objectives} describes in details the \textit{objective} dimension, \textit{features} are pointed out in subsection~\ref{sec:taxonomy_features} and machine learning algorithms are reported in subsection~\ref{sec:taxonomy_algorithms}.

\subsection{Malware Analysis Objectives} \label{sec:taxonomy_objectives}

Malware analysis, in general, demands for strong detection capabilities to find matches with the knowledge developed by investigating past samples. Anyway, the final goal of searching for those matches differs. For example, a malware analyst may be specifically interested in determining whether new suspicious samples are malicious or not, while another may be rather inspecting new malware looking for what family they likely belong to.
This subsection details the analysis goals of the surveyed papers, organized in three main objectives: malware detection (\S~\ref{sec:taxonomy_objective_malware_detection}), malware similarity analysis (\S~\ref{sec:taxonomy_objective_malware_similarity_analysis}) and malware category detection (\S~\ref{sec:taxonomy_objective_malware_category_detection}).

\subsubsection{Malware Detection} \label{sec:taxonomy_objective_malware_detection}

The most common objective in the context of malware analysis is detecting whether a given sample is malicious.
This objective is also the most important because knowing in advance that a sample is dangerous allows to block it before it becomes harmful. %
Indeed, the majority of reviewed works has this as main goal \cite{Shultz:2001,Kolter:2006,Ahmed2009,Chau2010,FirdausiLimErwinEtAl2010,AndersonQuistNeilEtAl2011,Santos:2011,AndersonStorlieLane2012,Yonts2012,SantosDevesaBrezoEtAl2013,EskandariKhorshidpourHashemi2013,Vadrevu2013,BaiWangZou2014,KruczkowskiSzynkiewicz2014,TamersoyRoundyChau2014,UppalSinhaMehraEtAl2014,Chen:2015,ElhadiMaarofBarry2015,FengXiongCaoEtAl2015,Ghiasi:2015,Ahmadi:2015,Kwon2015,Mao2015,Saxe2015,WuechnerOchoaPretschner2015,Raff2017}.
Depending on what machine learning technique is used, the generated output can be provided with a confidence value that
can be used by analysts to understand if a sample needs further inspection. %

\subsubsection{Malware Similarity Analysis} \label{sec:taxonomy_objective_malware_similarity_analysis}

Another relevant objective is spotting similarities among malware, for example to understand how novel samples differ from previous, known ones. 
We find four slightly different versions of this objective: \textit{variants detection}, \textit{families detection}, \textit{similarities detection} and \textit{differences detection}.

\paragraph{Variants Detection} %
Developing variants is one of the most effective and cheapest strategies for an attacker to evade detection mechanisms, while reusing as much as possible already available codes and resources.
Recognizing that a sample is actually a variant of a known malware prevents such strategy to succeed, and paves the way to understand how malware evolve over time through the development of new variants.
Also this objective has been deeply studied in literature, and several reviewed papers target the detection of variants.
Given a malicious sample $m$, variants detection consists in selecting from the available knowledge base the samples that are variants of $m$~\cite{%
	Gharacheh:2015,Ghiasi:2015,KhodamoradiFazlaliMardukhiEtAl2015,UpchurchZhou2015,LiangPangDai2016,Vadrevu2016}.
Considering the huge number of malicious samples received daily from major security firms, recognising variants of already known malware is crucial to reduce the workload for human analysts.

\paragraph{Families Detection} %
Given a malicious sample $m$, families detection consists in selecting from the available knowledge base the families that $m$ likely belongs to~\cite{LeeMody2006,HuangYeJiang2009,ParkReevesMulukutlaEtAl2010,Ye2010,DahlStokesDengEtAl2013,HuShinBhatkarEtAl2013,IslamTianBattenEtAl2013,Kong:2013,NariGhorbani2013,Ahmadi:2015,KawaguchiOmote2015,Lin:2015,MohaisenAlrawiMohaisen2015,Pai:2015,Raff2017}.
In this way, it is possible to associate unknown samples to already known families and, by consequence, provide an added-value information for further analyses.

\paragraph{Similarities Detection} %
Analysts can be interested in identifying the specific similarities and differences of the binaries to analyse with respect to those already analysed. %
Similarities detection consists in discovering what parts and aspects of a sample are similar to something that has been already examined in the past. It enables to focus on what is really new, and hence to discard the rest as it does not deserve further investigation~\cite{BaileyOberheideAndersenEtAl2007,BayerComparettiHlauschekEtAl2009,RieckTriniusWillemsEtAl2011,PalahanBabicChaudhuriEtAl2013,Egele:2014}.

\paragraph{Differences Detection} %
As a complement, also identifying what is different from everything else already observed in the past results worthwhile. As a matter of fact, differences can guide towards discovering novel aspects that should be analysed more in depth~\cite{BayerComparettiHlauschekEtAl2009,LindorferKolbitschComparetti2011,RieckTriniusWillemsEtAl2011,PalahanBabicChaudhuriEtAl2013,SantosBrezoUgarte-PedreroEtAl2013,Polino:2015}.

\subsubsection{Malware Category Detection} \label{sec:taxonomy_objective_malware_category_detection}

Malware can be categorized according to their prominent behaviours and objectives.
They can be interested in spying on users' activities and stealing their sensitive information (i.e., \textit{spyware}), encrypting documents and asking for a ransom (i.e., \textit{ransomware}), or gaining remote control of an infected machine (i.e., \textit{remote access toolkits}).
Using these categories is a coarse-grained yet significant way of describing malicious samples~\cite{Wong2006,Attaluri2009,ChenRoussopoulosLiangEtAl2012,Comar:2013,Kwon2015,Sexton2015}.
Although cyber security firms have not still agreed upon a standardized taxonomy of malware categories, effectively recognising the categories of a sample can add valuable information for the analysis.

\subsection{Malware Analysis Features} \label{sec:taxonomy_features}

This subsection deals with the features of samples that are considered for the analysis. How features are extracted from executables is reported in subsection~\ref{sec:taxonomy_feature_extraction}, while subsection~\ref{sec:taxonomy_pe_feature} details which specific features are taken into account.

\subsubsection{Feature Extraction} \label{sec:taxonomy_feature_extraction}

The information extraction process is performed through either static or dynamic analysis, or a combination of both, while examination and correlation are carried out by using machine learning techniques.
Approaches based on static analysis look at the content of samples without requiring their execution, while dynamic analysis works by running samples to examine their behaviour.
Several techniques can be used for dynamic malware analysis. \textit{Debuggers} are used for instruction level analysis. \textit{Simulators} model and show a behaviour similar to the environment expected by the malware, while \textit{emulators} replicate the behaviour of a system with higher accuracy but require more resources. Sandboxes are virtualised operating systems providing an isolated and reliable environment where to detonate malware. Refer to Ye \textit{et al.}~\cite{Ye2017} for a more detailed description of these techniques.
Execution traces are commonly used to extract features when dynamic analysis is employed. 
Reviewed articles generate execution traces by using either sandboxes~\cite{LeeMody2006,BayerComparettiHlauschekEtAl2009,Park2009,FirdausiLimErwinEtAl2010,AndersonQuistNeilEtAl2011,LindorferKolbitschComparetti2011,RieckTriniusWillemsEtAl2011,PalahanBabicChaudhuriEtAl2013,Graziano:2015,KawaguchiOmote2015,Lin:2015,Mao2015} or emulators~\cite{Asquith2015,LiangPangDai2016}. Also program analysis tools and techniques can be useful in the feature extraction process by providing, for example, disassembly code and control- and data-flow graphs.
An accurate disassembly code is important for obtaining correct \textit{Byte sequences} and~\textit{Opcodes} features (\S~\ref{sec:taxonomy_pe_feature}), while control- and data-flow graphs can be employed in the extraction of \textit{API and System Calls} (\S~\ref{sec:taxonomy_pe_feature}).
For an extensive dissertation on dynamic analyses, refer to~\cite{EST12}.

Among reviewed works, the majority relies on dynamic analyses~\cite{LeeMody2006,BaileyOberheideAndersenEtAl2007,BayerComparettiHlauschekEtAl2009,FirdausiLimErwinEtAl2010,ParkReevesMulukutlaEtAl2010,AndersonQuistNeilEtAl2011,LindorferKolbitschComparetti2011,RieckTriniusWillemsEtAl2011,%
Comar:2013,DahlStokesDengEtAl2013,NariGhorbani2013,PalahanBabicChaudhuriEtAl2013,KruczkowskiSzynkiewicz2014,UppalSinhaMehraEtAl2014,ElhadiMaarofBarry2015,Ghiasi:2015,KawaguchiOmote2015,Lin:2015,MohaisenAlrawiMohaisen2015,WuechnerOchoaPretschner2015,LiangPangDai2016}, while the others use, in equal proportions, either static analyses alone~\cite{Shultz:2001,Kolter:2006,Wong2006,Attaluri2009,Siddiqui:2009,Santos:2011,ChenRoussopoulosLiangEtAl2012,%
Yonts2012,HuShinBhatkarEtAl2013,Kong:2013,SantosBrezoUgarte-PedreroEtAl2013,Vadrevu2013,BaiWangZou2014,TamersoyRoundyChau2014,Ahmadi:2015,Caliskan-Islam:2015,Chen:2015,FengXiongCaoEtAl2015,Gharacheh:2015,KhodamoradiFazlaliMardukhiEtAl2015,Pai:2015,Sexton2015,SrakaewPiyanuntcharatsr:2015,UpchurchZhou2015} or a combination of static and dynamic techniques~\cite{JangBrumleyVenkataraman2011,AndersonStorlieLane2012%
,%
EskandariKhorshidpourHashemi2013,IslamTianBattenEtAl2013,SantosDevesaBrezoEtAl2013,Egele:2014,Graziano:2015,Polino:2015%
,Vadrevu2016}.
Depending on the specific features, extraction processes can be performed by applying either static, dynamic, or hybrid analysis.

\subsubsection{Portable Executable Features} \label{sec:taxonomy_pe_feature}

This section provides an overview on what features are used by reviewed papers to achieve the objectives outlined in section~\ref{sec:taxonomy_objectives}.
In many cases, surveyed works only refer to macro-classes without mentioning the specific features they employed.
As an example, when n-grams are used, only a minority of works mention the size of $n$.

\paragraph{Byte Sequences}
A binary can be characterised by computing features on its byte-level content.
Analysing the specific sequences of bytes in a PE is a widely employed static technique. %
A few works use chunks of bytes of specific sizes~\cite{Shultz:2001,SrakaewPiyanuntcharatsr:2015,Raff2017}, while many others rely on n-grams ~\cite{Kolter:2006,AndersonQuistNeilEtAl2011,JangBrumleyVenkataraman2011,RieckTriniusWillemsEtAl2011,AndersonStorlieLane2012,DahlStokesDengEtAl2013,UppalSinhaMehraEtAl2014,Ahmadi:2015,Chen:2015,FengXiongCaoEtAl2015,Lin:2015,Sexton2015,SrakaewPiyanuntcharatsr:2015,UpchurchZhou2015,WuechnerOchoaPretschner2015}.

An n-gram is a sequence of $n$ bytes, and features correspond to the different combination of these $n$ bytes, namely each feature represents how many times a specific combination of $n$ bytes occurs in the binary.
The majority of works that specified the size of used n-grams relies on sequences no longer than $3$ (i.e.  trigrams)~\cite{SrakaewPiyanuntcharatsr:2015,Lin:2015,Ahmadi:2015,AndersonQuistNeilEtAl2011,AndersonStorlieLane2012,Sexton2015,DahlStokesDengEtAl2013,IslamTianBattenEtAl2013}.
Indeed, the number of features to consider grows exponentially with $n$.

\paragraph{Opcodes}
Opcodes identify the machine-level operations executed by a PE, and can be extracted through static analyses by examining the assembly code~\cite{Wong2006,Attaluri2009,Ye2010,AndersonQuistNeilEtAl2011,AndersonStorlieLane2012,%
	HuShinBhatkarEtAl2013,Kong:2013,SantosBrezoUgarte-PedreroEtAl2013,SantosDevesaBrezoEtAl2013,Ahmadi:2015,Gharacheh:2015,KhodamoradiFazlaliMardukhiEtAl2015,Pai:2015,Sexton2015,SrakaewPiyanuntcharatsr:2015}.
Opcode frequency is one of the most commonly used feature. It measures the number of times each specific opcode appears within the assembly or is executed by a PE~\cite{Ye2010,KhodamoradiFazlaliMardukhiEtAl2015}. 
Others~\cite{AndersonStorlieLane2012,KhodamoradiFazlaliMardukhiEtAl2015} count opcode occurrences by aggregating them by operation type, e.g., mathematical instructions, memory access instructions.
Similarly to n-grams, also sequences of opcodes are used as features~\cite{Ye2010,Gharacheh:2015,KhodamoradiFazlaliMardukhiEtAl2015,SrakaewPiyanuntcharatsr:2015}. 

\paragraph{API and System Calls}
Similarly to opcodes, APIs and system calls enable the analysis of samples' behaviour, but at a higher level. %
They can be either extracted statically or dynamically by analysing the disassembly code (to get the list of all calls that can be potentially executed) or the execution traces (for the list of calls actually invoked).
While APIs allow to characterise what actions are executed by a sample~\cite{Ahmed2009,%
	IslamTianBattenEtAl2013,Kong:2013,BaiWangZou2014,Egele:2014,Ahmadi:2015,KawaguchiOmote2015,LiangPangDai2016}, looking at system call invocations provides a view on the interaction of the PE with the operating system~\cite{LeeMody2006,BayerComparettiHlauschekEtAl2009,ParkReevesMulukutlaEtAl2010,RieckTriniusWillemsEtAl2011,AndersonStorlieLane2012,%
	DahlStokesDengEtAl2013,PalahanBabicChaudhuriEtAl2013,SantosDevesaBrezoEtAl2013,Egele:2014,UppalSinhaMehraEtAl2014,Asquith2015,ElhadiMaarofBarry2015,Mao2015}.
Data extracted by observing APIs and system calls can be really large, and many works carry out additional processing to reduce feature space by using convenient data structures.
One of the most popular data structures to represent PE behaviour and extract program structure is the control flow graph.
This data structure allows compilers to produce an optimized version of the program itself and model control flow relationships~\cite{Allen:1970}.
Several works employ control flow graphs and their extensions for sample analysis, in combination with other feature classes~\cite{AndersonStorlieLane2012, EskandariKhorshidpourHashemi2013, Graziano:2015, Polino:2015, WuechnerOchoaPretschner2015}.

\paragraph{Network Activity}
A huge number of key information can be obtained by observing how the PE interacts with the network.
Contacted addresses and generated traffic can unveil valuable aspects, e.g. regarding the communication with a command and control centre.
Relevant features include statistics on used protocols, TCP/UDP ports, HTTP requests, DNS-level interactions. %
Many surveyed works require dynamic analysis to extract this kind of information~\cite{LeeMody2006,BaileyOberheideAndersenEtAl2007,BayerComparettiHlauschekEtAl2009,LindorferKolbitschComparetti2011,%
	NariGhorbani2013,Graziano:2015,Kwon2015,MohaisenAlrawiMohaisen2015,LiangPangDai2016}.
Other papers extract network-related inputs by monitoring the network and analysing incoming and outgoing traffic~\cite{Comar:2013,KruczkowskiSzynkiewicz2014,Vadrevu2016}.
A complementary approach consists in analysing download patterns of network users in a monitored network~\cite{Vadrevu2013}.
It does not require sample execution and focuses on network features related to the download of a sample, such as the website from which the file has been downloaded.

\paragraph{File System}
What file operations are executed by samples is fundamental to grasp evidence about the interaction with the environment and possibly detect attempts to gain persistence.
Features of interest mainly concern how many files are read or modified, what types of files and in what directories, and which files appear in not-infected/infected machines ~\cite{LeeMody2006,BaileyOberheideAndersenEtAl2007,Chau2010,%
	Kong:2013,Graziano:2015,Lin:2015,Mao2015,MohaisenAlrawiMohaisen2015}. %
Sandboxes and memory analysis toolkits include modules for monitoring interactions with the file system, usually modelled by counting the number of files created/deleted/modified by the PE. %
In~\cite{MohaisenAlrawiMohaisen2015}, the size of these files is considered as well, while Lin \textit{et al.} leverage the number of created hidden files~\cite{Lin:2015}.

A particularly relevant type of file system features are those extracted from the \textit{Windows Registry}.
The registry is one of the main sources of information for a PE about the environment, and also represents a fundamental tool to hook into the operating system, for example to gain persistence.
Discovering what keys are queried, created, deleted and modified can shed light on many significant characteristics of a sample~\cite{LeeMody2006,%
	Lin:2015,Mao2015,MohaisenAlrawiMohaisen2015}. %
Usually, works relying on file system inputs monitor also the Windows Registry.

\paragraph{CPU Registers}
The way CPU registers are used can also be a valuable indication, including whether any hidden register is used, and what values are stored in the registers, especially in the FLAGS register~\cite{Kong:2013, Ahmadi:2015, Egele:2014, Ghiasi:2015}. 

\paragraph{PE file characteristics}
A static analysis of a PE can provide a large set of valuable information such as sections, imports, symbols, used compilers~\cite{LeeMody2006, Yonts2012, Kirat2013, BaiWangZou2014, Asquith2015, Saxe2015}.

\paragraph{Strings}
A PE can be statically inspected to explicitly look for the strings it contains, such as code fragments, author signatures, file names, system resource information~\cite{Shultz:2001,IslamTianBattenEtAl2013,Saxe2015,Ahmadi:2015}.

\subsection{Malware Analysis Algorithms} \label{sec:taxonomy_algorithms}

This subsection reports what machine learning algorithms are used in surveyed works by organising them on the basis of whether the learning is supervised (\S~\ref{sec:taxonomy_algorithms_supervised}), unsupervised (\S~\ref{sec:taxonomy_algorithms_unsupervised}) or semi-supervised (\S~\ref{sec:taxonomy_algorithms_semi_supervised}).

\subsubsection{Supervised Learning} \label{sec:taxonomy_algorithms_supervised}

Supervised learning is the task of gaining knowledge by providing statistical models with correct instance examples, during a preliminary phase called training.
The supervised algorithms used by reviewed papers are 
\textit{rule-based classifier}~\cite{Shultz:2001, FengXiongCaoEtAl2015, Ghiasi:2015,Sexton2015,LiangPangDai2016, Tian:2008, LindorferKolbitschComparetti2011, Ahmed2009}, 
\textit{Bayes classifier}~\cite{SantosBrezoUgarte-PedreroEtAl2013,SantosDevesaBrezoEtAl2013,UppalSinhaMehraEtAl2014,KawaguchiOmote2015,WuechnerOchoaPretschner2015}, 
\textit{Na{\"i}ve Bayes}~\cite{Shultz:2001,Kolter:2006,FirdausiLimErwinEtAl2010,UppalSinhaMehraEtAl2014,KawaguchiOmote2015,Sexton2015,WuechnerOchoaPretschner2015},
\textit{Bayesian Network}~\cite{EskandariKhorshidpourHashemi2013,SantosBrezoUgarte-PedreroEtAl2013,SantosDevesaBrezoEtAl2013},
\textit{Support Vector Machine} (SVM)~\cite{Kolter:2006,Ahmed2009,FirdausiLimErwinEtAl2010,AndersonQuistNeilEtAl2011,ChenRoussopoulosLiangEtAl2012,Comar:2013,IslamTianBattenEtAl2013,Kong:2013,SantosBrezoUgarte-PedreroEtAl2013,SantosDevesaBrezoEtAl2013,KruczkowskiSzynkiewicz2014,UppalSinhaMehraEtAl2014,Ahmadi:2015,FengXiongCaoEtAl2015,KawaguchiOmote2015,Lin:2015,MohaisenAlrawiMohaisen2015,Sexton2015,WuechnerOchoaPretschner2015}, 
\textit{Multiple Kernel Learning}~\cite{AndersonStorlieLane2012},
\textit{Prototype-based Classification}~\cite{RieckTriniusWillemsEtAl2011},
\textit{Decision Tree}~\cite{Kolter:2006,Ahmed2009,FirdausiLimErwinEtAl2010,IslamTianBattenEtAl2013,NariGhorbani2013,SantosBrezoUgarte-PedreroEtAl2013,SantosDevesaBrezoEtAl2013,BaiWangZou2014,UppalSinhaMehraEtAl2014,KawaguchiOmote2015,KhodamoradiFazlaliMardukhiEtAl2015,MohaisenAlrawiMohaisen2015,SrakaewPiyanuntcharatsr:2015}, 
\textit{Random Forest}~\cite{Siddiqui:2009,Comar:2013,IslamTianBattenEtAl2013,UppalSinhaMehraEtAl2014,Ahmadi:2015,KawaguchiOmote2015,KhodamoradiFazlaliMardukhiEtAl2015,Kwon2015,Mao2015,WuechnerOchoaPretschner2015},
\textit{Gradient Boosting Decision Tree}~\cite{ChenRoussopoulosLiangEtAl2012,Sexton2015},
\textit{Logistic Model Tree}~\cite{Graziano:2015,DahlStokesDengEtAl2013,Sexton2015,PalahanBabicChaudhuriEtAl2013},
\textit{k-Nearest Neighbors} (k-NN)~\cite{LeeMody2006,Kong:2013,MohaisenAlrawiMohaisen2015,Raff2017,Ahmed2009,FirdausiLimErwinEtAl2010,IslamTianBattenEtAl2013,KawaguchiOmote2015,SantosBrezoUgarte-PedreroEtAl2013},
\textit{Artificial Neural Network}~\cite{DahlStokesDengEtAl2013,Saxe2015},
\textit{Multilayer Perceptron Neural Network}~\cite{FirdausiLimErwinEtAl2010}.

\subsubsection{Unsupervised Learning} \label{sec:taxonomy_algorithms_unsupervised}

Unsupervised approaches do not rely on any training phase and learn directly from unlabeled data.
Reviewed papers use these unsupervised learning algorithms: 
\textit{Clustering with locality sensitive hashing}~\cite{BayerComparettiHlauschekEtAl2009,TamersoyRoundyChau2014,UpchurchZhou2015},
\textit{Clustering with Distance and Similarity Metrics} (using either Euclidean~\cite{RieckTriniusWillemsEtAl2011,MohaisenAlrawiMohaisen2015} or Hamming distances~\cite{MohaisenAlrawiMohaisen2015}, or cosine~\cite{MohaisenAlrawiMohaisen2015} or Jaccard similarities~\cite{MohaisenAlrawiMohaisen2015,Polino:2015}),
\textit{Expectation Maximization}~\cite{Pai:2015}, 
\textit{k-Means Clustering}~\cite{HuangYeJiang2009,Pai:2015}, 
\textit{$k$-Medoids}~\cite{Ye2010},
\textit{Density-based Spatial Clustering of Applications with Noise}~\cite{Vadrevu2016},
\textit{Hierarchical Clustering}~\cite{JangBrumleyVenkataraman2011,MohaisenAlrawiMohaisen2015}, 
\textit{Prototype-based Clustering}~\cite{RieckTriniusWillemsEtAl2011},
\textit{Self-Organizing Maps}~\cite{ChenRoussopoulosLiangEtAl2012}.

\subsubsection{Semi-supervised Learning} \label{sec:taxonomy_algorithms_semi_supervised}

Semi-supervised learning combines both labeled and unlabeled data for feeding statistical models to acquire knowledge.
\textit{Learning with Local and Global Consistency} is used in~\cite{Santos:2011} while \textit{Belief Propagation} in~\cite{Chau2010, TamersoyRoundyChau2014, Chen:2015}.

\section{Characterization of Surveyed Papers} \label{sec:characterization}

In this section we %
characterize each reviewed paper on the basis of analysis objective, used machine learning algorithm and features.
Several details are also reported on the dataset used for the evaluation, including whether it is publicly available (\textit{Public} column), where samples have been collected from (\textit{Source} column) and whether the specific set of samples considered for the experiment is available (\textit{Available} column). Indeed, many works declare they do not use all the executables in the dataset but they do not specify what samples they choose, which prevents to reproduce their results. The \textit{Label} column states how samples have been labelled. Finally, \textit{Benign}, \textit{Malicious} and \textit{Total} columns report benign executables count, malware count and their sum, respectively.
\noindent \textbf{Malware detection}. Table~\ref{tab:char_malware_detection} lists all the reviewed works having malware detection as objective.
Most used features are byte sequences %
and API/system call invocations, derived by executing the samples.
Most of the works use more than one algorithm to find out the one guaranteeing more accurate results.
\noindent \textbf{Malware similarity analysis}. A table is provided for each version of this objective (\S~\ref{sec:taxonomy_objective_malware_similarity_analysis}). 
Tables~\ref{tab:char_malware_variants_selection} and~\ref{tab:char_malware_families_selection} describe the works dealing with variants detection and families detection, respectively. For both, APIs and system calls are largely used, as well as malware interactions with the environment, i.e. memory, file system, and CPU registers.
Tables~\ref{tab:char_malware_similarities_detection} and~\ref{tab:char_malware_differences_detection} report the papers on similarities and differences detection, respectively. All the analysed papers but~\cite{SantosBrezoUgarte-PedreroEtAl2013} rely on APIs and system calls collection. Works on differences detection, in general, do not take into account the interactions with the hosting system, while those on similarities detection do.

\noindent \textbf{Malware category detection}. These articles focus on the identification of specific threats and, thus, on particular features such as byte sequences, opcodes, function lengths and network activity.
Table~\ref{tab:char_malware_category_detection} reports the works whose objective is the detection of malware category.

\smallskip

By reasoning on what algorithms and features have been used and what have not for specific objectives, the provided characterisation allows to easily identify gaps in the literature and, thus, possible research directions to investigate. For instance, all works on \textit{differences detection} (see Table~\ref{tab:char_malware_differences_detection}) but~\cite{SantosBrezoUgarte-PedreroEtAl2013}, rely on dynamically extracted APIs and system calls for building their machine learning models. Novel approaches can be explored by taking into account other features that capture malware interactions with the environment (e.g., memory, file system, CPU registers and Windows Registry). %

\begin{landscape}
	\tiny
	\tiny
	\centering
	\begin{longtable}{| m{2cm} | m{2.5cm} | m{2.5cm} | m{3.5cm} | m{.5} | m{.5cm} | m{.5cm} | m{.5cm} | m{.5cm} | m{.5cm} | m{.5cm}}
		\caption{Characterization of surveyed papers having malware detection as objective.}
		\label{tab:char_malware_detection}
		\\ \hline
		\multirow{2}{*}{Paper} & \multirow{2}{*}{Algorithms} &  \multirow{2}{*}{Features} & \multirow{2}{*}{Limitations} &  \multicolumn{7}{c|}{Dataset samples} \\ \cline{5-11}
		&										   & 							 		   & 										  &		 
		\multicolumn{1}{c|}{Public} & \multicolumn{1}{c|}{Source} & \multicolumn{1}{c|}{Available} & \multicolumn{1}{c|}{Labeling} & \multicolumn{1}{c|}{Benign} & \multicolumn{1}{c|}{Malicious}  & \multicolumn{1}{c|}{Total}\\ \hline\hline
		\endfirsthead
		\caption{Characterization of surveyed papers having malware detection as objective. (Continued)}                              
		\\ \hline
		\multirow{2}{*}{Paper} & \multirow{2}{*}{Algorithms} &  \multirow{2}{*}{Features} & \multirow{2}{*}{Limitations} &  \multicolumn{7}{c|}{Dataset samples} \\ \cline{5-11}
		&										   & 							 		   & 										  &		 
		\multicolumn{1}{c|}{Public} & \multicolumn{1}{c|}{Source} & \multicolumn{1}{c|}{Available} & \multicolumn{1}{c|}{Labeling} & \multicolumn{1}{c|}{Benign} & \multicolumn{1}{c|}{Malicious} & \multicolumn{1}{c|}{Total}\\ \hline\hline
		\endhead
		Schultz \textit{et al}~\cite{Shultz:2001} & Rule-based classifier, Na{\"i}ve Bayes & Strings and byte sequences & Proposed solutions are not effective against encrypted executables and the dataset used in the evaluations is small. & \multicolumn{1}{c|}{$-$} & \multicolumn{1}{c|}{\makecell{FTP \\ sites}} & \multicolumn{1}{c|}{\checkmark} & \multicolumn{1}{c|}{Automated} & \multicolumn{1}{c|}{$1,001$} & \multicolumn{1}{c|}{$3,265$} & \multicolumn{1}{c|}{$4,266$}\\ \hline
		Kolter and \newline Maloof~\cite{Kolter:2006} & Decision Tree, Na{\"i}ve Bayes, SVM & Byte sequences & Payload classification fails in presence of binary obfuscation and the dataset used in the evaluations is very small. & \multicolumn{1}{c|}{$\xmark$} & \multicolumn{1}{c|}{\makecell{Internet, \\VX Heavens,\\and MITRE}} & \multicolumn{1}{c|}{\xmark} & \multicolumn{1}{c|}{Automated} & \multicolumn{1}{c|}{$1,971$} & \multicolumn{1}{c|}{$1,651$} & \multicolumn{1}{c|}{$3,622$}\\ \hline
		Ahmed \textit{et al.}~\cite{Ahmed2009} & Decision Tree, Na{\"i}ve Bayes, SVM & APIs/System calls & The dataset used in the experimental evaluations is very small. & \multicolumn{1}{c|}{$\checkmark$}  & \multicolumn{1}{c|}{\makecell{Legitimate apps\\and\\VX Heavens}} & \multicolumn{1}{c|}{\xmark} & \multicolumn{1}{c|}{$-$} & \multicolumn{1}{c|}{$100$} & \multicolumn{1}{c|}{$416$} & \multicolumn{1}{c|}{$516$}\\ \hline
		Chau \textit{et al.}~\cite{Chau2010} & Belief propagation & File system & Rare and new files cannot be accurately classified as benign or malicious. & \multicolumn{1}{c|}{$\checkmark$}  & \multicolumn{1}{c|}{\makecell{Symantec’s \\ Norton \\ Community \\ Watch}} & \multicolumn{1}{c|}{\xmark} & \multicolumn{1}{c|}{$-$} & \multicolumn{1}{c|}{$?$} & \multicolumn{1}{c|}{$?$} & \multicolumn{1}{c|}{$903 \cdot 10^6$}\\ \hline
		Firdausi \textit{et al.}~\cite{FirdausiLimErwinEtAl2010} & Decision Tree, Na{\"i}ve Bayes, SVM, \textit{k}-NN, Multilayer Perceptron Neural Network & APIs/System calls, file system, and Windows Registry & The dataset used in the experimental evaluations is very small. & \multicolumn{1}{c|}{$\checkmark$}  & \multicolumn{1}{c|}{\makecell{Windows\\XP SP2}} & \multicolumn{1}{c|}{\xmark} & \multicolumn{1}{c|}{$-$} & \multicolumn{1}{c|}{$250$} & \multicolumn{1}{c|}{$220$} & \multicolumn{1}{c|}{$470$}\\ \hline
		Anderson \textit{et al.}~\cite{AndersonQuistNeilEtAl2011} & SVM & Byte sequences and APIs/system calls & The dataset used in the evaluations is small. & \multicolumn{1}{c|}{$-$}  & \multicolumn{1}{c|}{\makecell{$-$}} & \multicolumn{1}{c|}{\xmark} & \multicolumn{1}{c|}{$-$} & \multicolumn{1}{c|}{$615$} & \multicolumn{1}{c|}{$1,615$} & \multicolumn{1}{c|}{$2,230$}\\ \hline
		Santos \textit{et al.}~\cite{Santos:2011} & Learning with Local and Global Consistency & Byte sequences & Proposed approach is not effective against packed malware and requires manual labeling of a portion of the small dataset. In particular, the dataset used in the experimental evaluations is small. & \multicolumn{1}{c|}{$\checkmark$}  & \multicolumn{1}{c|}{\makecell{Own\\machines\\and\\ VX Heavens}} & \multicolumn{1}{c|}{\xmark} & \multicolumn{1}{c|}{$-$} & \multicolumn{1}{c|}{$1,000$} & \multicolumn{1}{c|}{$1,000$} & \multicolumn{1}{c|}{$2,000$}\\ \hline
		Anderson \textit{et al.}~\cite{AndersonStorlieLane2012} & Multiple Kernel Learning & Byte sequences, opcodes, and APIs/system calls & Instruction categorization is not optimal. %
		& \multicolumn{1}{c|}{$\xmark$}  & \multicolumn{1}{c|}{\makecell{Offensive\\Computing}} & \multicolumn{1}{c|}{\xmark} & \multicolumn{1}{c|}{$-$}  & \multicolumn{1}{c|}{$776$} & \multicolumn{1}{c|}{$21,716$} & \multicolumn{1}{c|}{$22,492$}\\ \hline
		Yonts \cite{Yonts2012} & Rule-based classifier & PE file characteristics & Only a subset of all the potential low level attributes is considered. & \multicolumn{1}{c|}{$\xmark$}  & \multicolumn{1}{c|}{\makecell{SANS\\Institute}} & \multicolumn{1}{c|}{\xmark} & \multicolumn{1}{c|}{$-$} & \multicolumn{1}{c|}{$65,000$} & \multicolumn{1}{c|}{$25 \cdot 10^5$} & \multicolumn{1}{c|}{$25.65 \cdot 10^5$}\\ \hline
		\multicolumn{11}{r}{\textit{Continue on the next page}} \\ 
		Eskandari \textit{et al.}~\cite{EskandariKhorshidpourHashemi2013} & Bayesian Network & APIs/System calls & Ignore specific instructions. Evasion/obfuscation techniques and samples requiring user interactions reduce the effectiveness of the proposed approach. The dataset is small. & \multicolumn{1}{c|}{$\xmark$}  & \multicolumn{1}{c|}{\makecell{Research\\Laboratory at\\Shiraz\\University}} & \multicolumn{1}{c|}{\xmark} & \multicolumn{1}{c|}{$-$} & \multicolumn{1}{c|}{$1,000$} & \multicolumn{1}{c|}{$2,000$} & \multicolumn{1}{c|}{$3,000$} \\ \hline
		Santos \textit{et al.}~\cite{SantosDevesaBrezoEtAl2013} & Bayesian Network, Decision Tree, \textit{k}-NN, SVM & Opcodes, APIs/system calls, and raised exceptions & Packed malware, evasion techniques, and samples requiring user interactions reduce the accuracy of the proposed solution. The dataset is small. & \multicolumn{1}{c|}{$\checkmark$}  & \multicolumn{1}{c|}{\makecell{Own\\machines\\and\\VX Heavens}} & \multicolumn{1}{c|}{\xmark} & \multicolumn{1}{c|}{$-$} & \multicolumn{1}{c|}{$1,000$} & \multicolumn{1}{c|}{$1,000$} & \multicolumn{1}{c|}{$2,000$}\\ \hline
		Vadrevu \textit{et al.}~\cite{Vadrevu2013} & Random Forest & PE file characteristics and network & Requires a huge number of samples labeled either as malicious or benign. & \multicolumn{1}{c|}{$\xmark$}  & \multicolumn{1}{c|}{\makecell{Georgia\\Institute of\\Technology}} & \multicolumn{1}{c|}{\xmark} & \multicolumn{1}{c|}{Automated} & \multicolumn{1}{c|}{$170,780$} & \multicolumn{1}{c|}{$15,182$} & \multicolumn{1}{c|}{$185,962$}\\ \hline
		Bai \textit{et al.}~\cite{BaiWangZou2014} & Decision Tree, Random Forest & PE file characteristics & Assume that samples are not packed and malware authors can properly modify PE header to remain undetected. & \multicolumn{1}{c|}{$\checkmark$}  & \multicolumn{1}{c|}{\makecell{Windows and \\Program Files \\folders and\\ VXHeavens}} & \multicolumn{1}{c|}{\xmark} & \multicolumn{1}{c|}{Automated} & \multicolumn{1}{c|}{$8,592$} & \multicolumn{1}{c|}{$10,521$} & \multicolumn{1}{c|}{$19,113$}\\ \hline
		Kruczkowski and Szynkiewicz~\cite{KruczkowskiSzynkiewicz2014} & SVM & Network & The dataset used in the experimental evaluations is small. & \multicolumn{1}{c|}{$\xmark$}  & \multicolumn{1}{c|}{\makecell{N6 Platform}} & \multicolumn{1}{c|}{\xmark} & \multicolumn{1}{c|}{$-$} & \multicolumn{1}{c|}{$?$} & \multicolumn{1}{c|}{$?$} & \multicolumn{1}{c|}{$1,015$}\\ \hline
		Tamersoy \textit{et al.}~\cite{TamersoyRoundyChau2014} & Clustering with locality sensitive hashing & File system & Rare and new files cannot be accurately classified as benign or malicious. & \multicolumn{1}{c|}{$\checkmark$}  & \multicolumn{1}{c|}{\makecell{Symantec’s\\Norton\\Community\\Watch}} & \multicolumn{1}{c|}{\xmark} & \multicolumn{1}{c|}{$-$}  & \multicolumn{1}{c|}{$1,663,506$} & \multicolumn{1}{c|}{$47,956$} & \multicolumn{1}{c|}{$4,970,865$}\\ \hline
		Uppal \textit{et al.}~\cite{UppalSinhaMehraEtAl2014} & Decision Tree, Random Forest, Na{\"i}ve Bayes, SVM & Byte sequences and APIs/system calls & The dataset used in the experimental evaluations is very small.  & \multicolumn{1}{c|}{$\checkmark$}  & \multicolumn{1}{c|}{\makecell{Legitimate apps\\and\\VX Heavens}} & \multicolumn{1}{c|}{\xmark} & \multicolumn{1}{c|}{$-$}   & \multicolumn{1}{c|}{$150$} & \multicolumn{1}{c|}{$120$} & \multicolumn{1}{c|}{$270$}\\ \hline
		Chen \textit{et al.}~\cite{Chen:2015} & Belief propagation & File system & Rare and new files cannot be accurately classified as benign or malicious. & \multicolumn{1}{c|}{$\xmark$}  & \multicolumn{1}{c|}{\makecell{Comodo Cloud\\Security Center}} & \multicolumn{1}{c|}{\xmark} & \multicolumn{1}{c|}{$-$} & \multicolumn{1}{c|}{$19,142$} & \multicolumn{1}{c|}{$2,883$} & \multicolumn{1}{c|}{$69,165$}\\ \hline
		Elhadi \textit{et al.}~\cite{ElhadiMaarofBarry2015} & Malicious graph matching & APIs/System calls &  The dataset used in the experimental evaluations is extremely small.& \multicolumn{1}{c|}{$\checkmark$}  & \multicolumn{1}{c|}{\makecell{VX Heavens}} & \multicolumn{1}{c|}{\checkmark} & \multicolumn{1}{c|}{$-$} & \multicolumn{1}{c|}{$10$} & \multicolumn{1}{c|}{$75$} & \multicolumn{1}{c|}{$85$}\\ \hline
		Feng \textit{et al.}~\cite{FengXiongCaoEtAl2015} & Rule-based classifier, SVM & Byte sequences & Only specific malware classes are considered for the approach evaluation.& \multicolumn{1}{c|}{$\xmark$}  & \multicolumn{1}{c|}{\makecell{Windows system\\files and own\\AV platform}} & \multicolumn{1}{c|}{\xmark} & \multicolumn{1}{c|}{$-$}  & \multicolumn{1}{c|}{$100,000$} & \multicolumn{1}{c|}{$135,064$} & \multicolumn{1}{c|}{$235,064$} \\ \hline
		Ghiasi \textit{et al.}~\cite{Ghiasi:2015} & Rule-based classifier & APIs/System calls and CPU registers & APIs/System calls categorization could be not optimal and the dataset size is small. & \multicolumn{1}{c|}{$\checkmark$}  & \multicolumn{1}{c|}{\makecell{Windows XP \\system and\\Program Files\\folders and\\private repository}} & \multicolumn{1}{c|}{\xmark} & \multicolumn{1}{c|}{$-$} & \multicolumn{1}{c|}{$390$} & \multicolumn{1}{c|}{$850$} & \multicolumn{1}{c|}{$1,240$}\\ \hline
		Kwon \textit{et al.}~\cite{Kwon2015} & Random Forest & Network & Not able to detect bots with rootkit capabilities. & \multicolumn{1}{c|}{$\checkmark$}  & \multicolumn{1}{c|}{\makecell{Symantec’s\\Worldwide\\ Intelligence\\Network \\Environment}} & \multicolumn{1}{c|}{\xmark} & \multicolumn{1}{c|}{$-$}  & \multicolumn{1}{c|}{$?$} & \multicolumn{1}{c|}{$?$} & \multicolumn{1}{c|}{$24 * 10^6$}\\ \hline
		Mao \textit{et al.}~\cite{Mao2015} & Random Forest & APIs/System calls, file system, and Windows Registry & Evasion techniques and samples requiring user interactions reduce the accuracy of the proposed approach. The dataset is small. & \multicolumn{1}{c|}{$\checkmark$}  & \multicolumn{1}{c|}{\makecell{Windows\\XP SP3 and\\VX Heavens}} & \multicolumn{1}{c|}{\xmark} & \multicolumn{1}{c|}{$-$} & \multicolumn{1}{c|}{$534$} & \multicolumn{1}{c|}{$7,257$} & \multicolumn{1}{c|}{$7,791$}\\ \hline
		Saxe and Berlin~\cite{Saxe2015} & Neural Networks & Strings and PE file characteristics & Label assigned to training set may be inaccurate and the accuracy of the proposed approach decreases substantially when samples are obfuscated. & \multicolumn{1}{c|}{$\xmark$}  & \multicolumn{1}{c|}{\makecell{Legitimate apps\\and own malware\\database}} & \multicolumn{1}{c|}{\xmark} & \multicolumn{1}{c|}{Automated} &  \multicolumn{1}{c|}{$81,910$} & \multicolumn{1}{c|}{$350,016$}  & \multicolumn{1}{c|}{$431,926$}\\ \hline
		Srakaew \textit{et al.}~\cite{SrakaewPiyanuntcharatsr:2015} & Decision Tree & Byte sequences and opcodes & Obfuscation techniques reduce detection accuracy. & \multicolumn{1}{c|}{$\xmark$}  & \multicolumn{1}{c|}{\makecell{Legitimate files\\and apps and\\CWSandbox}} & \multicolumn{1}{c|}{\xmark} & \multicolumn{1}{c|}{$-$}  & \multicolumn{1}{c|}{$600$} & \multicolumn{1}{c|}{$3,851$} & \multicolumn{1}{c|}{$69,165$}\\ \hline
		W\"{u}chner \textit{et al.}~\cite{WuechnerOchoaPretschner2015} & Na{\"i}ve Bayes, Random Forest, SVM & Byte sequences, APIs/system calls, file system, and Windows Registry & Obfuscation techniques applied by the authors may not reflect the ones of real-world samples. The dataset is small. & \multicolumn{1}{c|}{$\xmark$}  & \multicolumn{1}{c|}{\makecell{Legitimate app\\downloads and\\Malicia}} & \multicolumn{1}{c|}{\xmark} & \multicolumn{1}{c|}{$-$}& \multicolumn{1}{c|}{$513$} & \multicolumn{1}{c|}{$6,994$} & \multicolumn{1}{c|}{$7,507$}\\ \hline
		Raff and\newline Nicholas~\cite{Raff2017} & $k$-NN with Lempel-Ziv Jaccard distance & Byte sequences & Obfuscation techniques reduce detection accuracy.& \multicolumn{1}{c|}{$\xmark$}  & \multicolumn{1}{c|}{\makecell{Industry partner}} & \multicolumn{1}{c|}{\xmark} & \multicolumn{1}{c|}{$-$}& \multicolumn{1}{c|}{$240,000$} & \multicolumn{1}{c|}{$237,349$} & \multicolumn{1}{c|}{$477,349$}\\ \hline
	\end{longtable}
\end{landscape}
\normalsize

\begin{landscape}
	\scriptsize
	\centering
	\begin{longtable}{| m{2.5cm} | m{2.2cm} | m{2.5cm} | m{3cm} | m{.5} | m{.5cm} | m{.5cm} | m{.5cm} | m{.5cm} | m{.5cm} | m{.5cm}}
		\caption{Characterization of surveyed papers having malware variants selection as objective.
			$^{1}$Instead of using machine learning techniques, Gharacheh \textit{et al.} rely on Hidden Markov Models to detect variants of the same malicious sample~\cite{Gharacheh:2015}. } %
		\\ \hline
		\multirow{2}{*}{Paper} & \multirow{2}{*}{Algorithms} &  \multirow{2}{*}{Features} & \multirow{2}{*}{Limitations} &  \multicolumn{7}{c|}{Dataset samples} \\ \cline{5-11}
		&										   & 							 		   & 										  &		 
		\multicolumn{1}{c|}{Public} & \multicolumn{1}{c|}{Source} & \multicolumn{1}{c|}{Available} & \multicolumn{1}{c|}{Labeling} & \multicolumn{1}{c|}{Benign} & \multicolumn{1}{c|}{Malicious}  & \multicolumn{1}{c|}{Total}\\ \hline\hline
		\endfirsthead
		\caption{Characterization of surveyed papers having malware detection as objective. (Continued)}                              
		\\ \hline
		\multirow{2}{*}{Paper} & \multirow{2}{*}{Algorithms} &  \multirow{2}{*}{Features} & \multirow{2}{*}{Limitations} &  \multicolumn{7}{c|}{Dataset samples} \\ \cline{5-11}
		&										   & 							 		   & 										  &		 
		\multicolumn{1}{c|}{Public} & \multicolumn{1}{c|}{Source} & \multicolumn{1}{c|}{Available} & \multicolumn{1}{c|}{Labeling} & \multicolumn{1}{c|}{Benign} & \multicolumn{1}{c|}{Malicious} & \multicolumn{1}{c|}{Total}\\ \hline\hline
		\endhead
		\label{tab:char_malware_variants_selection}

		Gharacheh \newline\textit{et al.}~\cite{Gharacheh:2015} & -\footnotemark & Opcodes & Opcode sequence is not optimal and the dataset size is very small. & \multicolumn{1}{c|}{$\checkmark$}  & \multicolumn{1}{c|}{\makecell{Cygwin and\\ VX Heavens}} & \multicolumn{1}{c|}{\xmark} & \multicolumn{1}{c|}{$-$} & \multicolumn{1}{c|}{$?$} & \multicolumn{1}{c|}{$?$} & \multicolumn{1}{c|}{$740$}\\ \hline
		Ghiasi \textit{et al.}~\cite{Ghiasi:2015} & Rule-based classifier & APIs/System calls and CPU registers & APIs/System calls categorization could be not optimal and the dataset size is small.& \multicolumn{1}{c|}{$\checkmark$}  & \multicolumn{1}{c|}{\makecell{\makecell{Windows XP \\system and\\Program Files\\folders and\\private repository}}} & \multicolumn{1}{c|}{\xmark} & \multicolumn{1}{c|}{$-$} & \multicolumn{1}{c|}{$390$} & \multicolumn{1}{c|}{$850$} & \multicolumn{1}{c|}{$1,240$}\\ \hline
		Khodamoradi \newline\textit{et al.}~\cite{KhodamoradiFazlaliMardukhiEtAl2015} & Decision Tree, Random Forest & Opcodes & Opcode sequence is not optimal and the dataset size is very small. & \multicolumn{1}{c|}{$\checkmark$}  & \multicolumn{1}{c|}{\makecell{Windows XP \\system and\\Program Files\\folders and\\self-generated\\metamorphic\\malware}} & \multicolumn{1}{c|}{\xmark} & \multicolumn{1}{c|}{$-$}& \multicolumn{1}{c|}{$550$} & \multicolumn{1}{c|}{$280$} & \multicolumn{1}{c|}{$830$}\\ \hline
		Upchurch and\newline Zhou~\cite{UpchurchZhou2015} & Clustering with locality sensitive hashing & Byte sequences & The dataset size is extremely small.& \multicolumn{1}{c|}{$\xmark$}  & \multicolumn{1}{c|}{\makecell{Sampled from\\security\\incidents}} & \multicolumn{1}{c|}{\checkmark} & \multicolumn{1}{c|}{Manual}& \multicolumn{1}{c|}{$0$} & \multicolumn{1}{c|}{$85$} & \multicolumn{1}{c|}{$85$}\\ \hline
		Liang \textit{et al.}~\cite{LiangPangDai2016} & Rule-based classifier & APIs/System calls, file system, Windows Registry, and network & Monitored API/system call set could be not optimal and the dataset size is small. & \multicolumn{1}{c|}{$\xmark$}  & \multicolumn{1}{c|}{\makecell{Anubis website}} & \multicolumn{1}{c|}{\xmark} & \multicolumn{1}{c|}{$-$}& \multicolumn{1}{c|}{$0$} & \multicolumn{1}{c|}{$330,248$} & \multicolumn{1}{c|}{$330,248$}\\ \hline
		\multicolumn{11}{r}{\textit{Continue on the next page}} \\ 
		Vadrevu and \newline Perdisci~\cite{Vadrevu2016} & DBSCAN clustering & APIs/System calls, PE file characteristics, and network & Evasion techniques and samples requiring user interactions reduce the accuracy of the proposed approach.& \multicolumn{1}{c|}{$\xmark$}  & \multicolumn{1}{c|}{\makecell{Security company\\and large\\Research\\Institute}} & \multicolumn{1}{c|}{\xmark} & \multicolumn{1}{c|}{$-$}& \multicolumn{1}{c|}{$0$} & \multicolumn{1}{c|}{$1,651,906$} & \multicolumn{1}{c|}{$1,651,906$}\\ \hline
	\end{longtable}

	\tiny
	
	\begin{longtable}{| m{2.5cm} | m{2.2cm} | m{2.5cm} | m{3cm} | m{1} | m{.8cm} | m{1cm} | m{1cm} | m{1cm} | m{1cm} | m{1cm}}
		\caption{Characterization of surveyed papers having malware families selection as objective. $^{2}$Asquith describes aggregation overlay graphs for storing PE metadata, without further discussing any machine learning technique that could be applied on top of these new data structures.}%
		\label{tab:char_malware_families_selection}
		\\ \hline
		\multirow{2}{*}{Paper} & \multirow{2}{*}{Algorithms} &  \multirow{2}{*}{Features} & \multirow{2}{*}{Limitations} &  \multicolumn{7}{c|}{Dataset samples} \\ \cline{5-11}
		&										   & 							 		   & 										  &		 
		\multicolumn{1}{c|}{Public} & \multicolumn{1}{c|}{Source} & \multicolumn{1}{c|}{Available} & \multicolumn{1}{c|}{Labeling} & \multicolumn{1}{c|}{Benign} & \multicolumn{1}{c|}{Malicious}  & \multicolumn{1}{c|}{Total}\\ \hline\hline
		\endfirsthead
		\caption{Characterization of surveyed papers having malware families selection as objective. (Continued)}                              
		\\ \hline
		\multirow{2}{*}{Paper} & \multirow{2}{*}{Algorithms} &  \multirow{2}{*}{Features} & \multirow{2}{*}{Limitations} &  \multicolumn{7}{c|}{Dataset samples} \\ \cline{5-11}
		&										   & 							 		   & 										  &		 
		\multicolumn{1}{c|}{Public} & \multicolumn{1}{c|}{Source} & \multicolumn{1}{c|}{Available} & \multicolumn{1}{c|}{Labeling} & \multicolumn{1}{c|}{Benign} & \multicolumn{1}{c|}{Malicious}  & \multicolumn{1}{c|}{Total}\\ \hline\hline
		\endhead
		Huang \textit{et al.}~\cite{HuangYeJiang2009} & \textit{k}-Means-like algorithm & Byte sequences & Instruction sequence categorization could be not optimal and the dataset size is small.  & \multicolumn{1}{c|}{$\xmark$}  & \multicolumn{1}{c|}{\makecell{Kingsoft Corporation}} & \multicolumn{1}{c|}{\xmark} & \multicolumn{1}{c|}{$-$} & \multicolumn{1}{c|}{$0$} & \multicolumn{1}{c|}{$2,029$} & \multicolumn{1}{c|}{$2,029$}\\ \hline
		Park \textit{et al.}~\cite{ParkReevesMulukutlaEtAl2010} & Malicious graph matching & APIs/System calls & Approach vulnerable to APIs/system calls injection and the dataset used in the experimental evaluations is very small. & \multicolumn{1}{c|}{$\xmark$}  & \multicolumn{1}{c|}{\makecell{Legitimate apps and\\Anubis Sandbox}} & \multicolumn{1}{c|}{\xmark} & \multicolumn{1}{c|}{Automated} & \multicolumn{1}{c|}{$80$} & \multicolumn{1}{c|}{$300$} & \multicolumn{1}{c|}{$380$}\\ \hline
		Ye \textit{et al.}~\cite{Ye2010} & $k$-Medoids variants & Opcodes & Instruction categorization could be not optimal. & \multicolumn{1}{c|}{$\xmark$}  & \multicolumn{1}{c|}{\makecell{Kingsoft Corporation}} & \multicolumn{1}{c|}{\xmark} & \multicolumn{1}{c|}{$-$} & \multicolumn{1}{c|}{$0$} & \multicolumn{1}{c|}{$11,713$} & \multicolumn{1}{c|}{$11,713$}\\ \hline
		Dahl \textit{et al.}~\cite{DahlStokesDengEtAl2013} & Logistic Regression, Neural Networks & Byte sequences and APIs/system calls & The authors obtain a high two-class error rate.  & \multicolumn{1}{c|}{$\xmark$}  & \multicolumn{1}{c|}{\makecell{Microsoft}} & \multicolumn{1}{c|}{\xmark} & \multicolumn{1}{c|}{\makecell{Mostly manual}} & \multicolumn{1}{c|}{$817,485$} & \multicolumn{1}{c|}{$1,843,359$} & \multicolumn{1}{c|}{$3,760,844$}\\ \hline
		Hu \textit{et al.}~\cite{HuShinBhatkarEtAl2013} & Prototype-based clustering & Opcodes & Obfuscation techniques reduce the effectiveness of their prototype for malware family selection. & \multicolumn{1}{c|}{$\xmark$}  & \multicolumn{1}{c|}{$-$} & \multicolumn{1}{c|}{\xmark} & \multicolumn{1}{c|}{\makecell{Manual\\and\\automated}} & \multicolumn{1}{c|}{$0$} & \multicolumn{1}{c|}{$137,055$} & \multicolumn{1}{c|}{$137,055$}\\ \hline
		\multicolumn{11}{r}{\textit{Continue on the next page}} \\ 
		Islam \textit{et al.}~\cite{IslamTianBattenEtAl2013} & Decision Tree, \textit{k}-NN, Random Forest, SVM & Strings, byte sequences and APIs/system calls & The proposed approach is less effective novel samples. The dataset is small. & \multicolumn{1}{c|}{$\xmark$}  & \multicolumn{1}{c|}{\makecell{CA Labs}} & \multicolumn{1}{c|}{\xmark} & \multicolumn{1}{c|}{$-$} & \multicolumn{1}{c|}{$51$} & \multicolumn{1}{c|}{$2,398$} & \multicolumn{1}{c|}{$2,939$}\\ \hline
		Kong and\newline Yan~\cite{Kong:2013} & SVM, \textit{k}-NN & Opcodes, memory, file system, and CPU registers & Significant differences in samples belonging to the same family reduce the proposed approach accuracy. & \multicolumn{1}{c|}{$\xmark$}  & \multicolumn{1}{c|}{\makecell{Offensive Computing}} & \multicolumn{1}{c|}{\xmark} & \multicolumn{1}{c|}{Automated} & \multicolumn{1}{c|}{$0$} & \multicolumn{1}{c|}{$526,179$} & \multicolumn{1}{c|}{$526,179$}\\ \hline
		Nari and Ghorbani~\cite{NariGhorbani2013} & Decision Tree & Network & Network features are extracted by a commercial traffic analyzer.  The dataset used in the experimental evaluations is small. & \multicolumn{1}{c|}{$\xmark$}  & \multicolumn{1}{c|}{\makecell{Communication\\Research Center\\Canada}} & \multicolumn{1}{c|}{\xmark} & \multicolumn{1}{c|}{Automated} & \multicolumn{1}{c|}{$0$} & \multicolumn{1}{c|}{$3,768$} & \multicolumn{1}{c|}{$3,768$} \\ \hline
		Ahmadi \textit{et al.}~\cite{Ahmadi:2015}& SVM, Random Forest, Gradient Boosting Decision Tree & Byte sequences, opcodes, APIs/system calls, Windows Registry, CPU registers, and PE file characteristics & Selected features can be further reduced to have a clearer view of the reasons behind sample classification.  & \multicolumn{1}{c|}{$\checkmark$}  & \multicolumn{1}{c|}{\makecell{Microsoft's malware\\classification\\challenge}} & \multicolumn{1}{c|}{\xmark} & \multicolumn{1}{c|}{$-$} & \multicolumn{1}{c|}{$0$} & \multicolumn{1}{c|}{$21,741$} & \multicolumn{1}{c|}{$21,741$}\\ \hline
		Asquith~\cite{Asquith2015}& -\footnotemark & APIs/System calls, memory, file system, PE file characteristics, and raised exceptions &  -$^5$  & \multicolumn{1}{c|}{$-$}  & \multicolumn{1}{c|}{$-$} & \multicolumn{1}{c|}{$-$} & \multicolumn{1}{c|}{$-$} & \multicolumn{1}{c|}{$-$} & \multicolumn{1}{c|}{$-$} & \multicolumn{1}{c|}{$-$}\\ \hline
		Lin \textit{et al.}~\cite{Lin:2015} & SVM & Byte sequences, APIs/system calls, file system, and CPU registers & Selected API/system call set could be not optimal. Evasion techniques and samples requiring user interactions reduce the accuracy of the proposed approach.   The dataset is small. & \multicolumn{1}{c|}{$\xmark$}  & \multicolumn{1}{c|}{\makecell{Own sandbox}} & \multicolumn{1}{c|}{\xmark} & \multicolumn{1}{c|}{$-$} & \multicolumn{1}{c|}{$389$} & \multicolumn{1}{c|}{$3,899$} & \multicolumn{1}{c|}{$4,288$}\\ \hline
		\multicolumn{11}{r}{\textit{Continue on the next page}} \\ 
		Kawaguchi and \newline Omote~\cite{KawaguchiOmote2015} & Decision Tree, Random Forest, \textit{k}-NN, Na{\"i}ve Bayes & APIs/System calls & This classification approach can be easily evaded by real-world malware. The dataset used in the experimental evaluations is very small. & \multicolumn{1}{c|}{$\xmark$}  & \multicolumn{1}{c|}{\makecell{FFRI Inc.}} & \multicolumn{1}{c|}{\xmark} & \multicolumn{1}{c|}{$-$} & \multicolumn{1}{c|}{$236$} & \multicolumn{1}{c|}{$408$} & \multicolumn{1}{c|}{$644$}\\ \hline
		Mohaisen \textit{et al.}~\cite{MohaisenAlrawiMohaisen2015} & Decision Tree, \textit{k}-NN, SVM, Clustering with with different similarity measures, Hierarchical clustering & File system, Windows Registry, CPU registers, and network & Evasion techniques reduce the accuracy of the proposed solution.  & \multicolumn{1}{c|}{$\xmark$}  & \multicolumn{1}{c|}{\makecell{AMAL system}} & \multicolumn{1}{c|}{\xmark} & \multicolumn{1}{c|}{\makecell{Manual\\and\\automated}} & \multicolumn{1}{c|}{$0$} & \multicolumn{1}{c|}{$115,157$} & \multicolumn{1}{c|}{$115,157$}\\ \hline
		Pai \textit{et al.}~\cite{Pai:2015} & \textit{k}-Means, Expectation Maximization & Opcodes & Obfuscation techniques reduce the effectiveness of the employed approach. The dataset is small.  & \multicolumn{1}{c|}{$\xmark$}  & \multicolumn{1}{c|}{\makecell{Cygwin utility\\files and Malicia}} & \multicolumn{1}{c|}{\xmark} & \multicolumn{1}{c|}{$-$} & \multicolumn{1}{c|}{$213$} & \multicolumn{1}{c|}{$8,052$} & \multicolumn{1}{c|}{$8,265$}\\ \hline
		Raff and\newline Nicholas~\cite{Raff2017} & $k$-NN with Lempel-Ziv Jaccard distance & Byte sequences & Obfuscation techniques reduce detection accuracy.& \multicolumn{1}{c|}{$\xmark$}  & \multicolumn{1}{c|}{\makecell{Industry partner}} & \multicolumn{1}{c|}{\xmark} & \multicolumn{1}{c|}{$-$}& \multicolumn{1}{c|}{$240,000$} & \multicolumn{1}{c|}{$237,349$} & \multicolumn{1}{c|}{$477,349$}\\ \hline
	\end{longtable}
		\scriptsize
		\centering
		\begin{longtable}{| m{2.5cm} | m{2.7cm} | m{2.5cm} | m{2.9cm} | m{.5} | m{.5cm} | m{.5cm} | m{.5cm} | m{.5cm} | m{.5cm} | m{.5cm}}
			\caption{Characterization of surveyed papers having malware similarities detection as objective. $^{3}$SVM is used only for computing the optimal values of weight factors associated to each feature chosen to detect similarities among malicious samples.}
			\label{tab:char_malware_similarities_detection}
			\\ \hline
			\multirow{2}{*}{Paper} & \multirow{2}{*}{Algorithms} &  \multirow{2}{*}{Features} & \multirow{2}{*}{Limitations} &  \multicolumn{7}{c|}{Dataset samples} \\ \cline{5-11}
			&										   & 							 		   & 										  &		 
			\multicolumn{1}{c|}{Public} & \multicolumn{1}{c|}{Source} & \multicolumn{1}{c|}{Available} & \multicolumn{1}{c|}{Labeling} & \multicolumn{1}{c|}{Benign} & \multicolumn{1}{c|}{Malicious}  & \multicolumn{1}{c|}{Total}\\ \hline\hline
			\endfirsthead
			\caption{Characterization of surveyed papers having malware similarities detection as objective. (Continued)} %
			\\ \hline
			\multirow{2}{*}{Paper} & \multirow{2}{*}{Algorithms} &  \multirow{2}{*}{Features} & \multirow{2}{*}{Limitations} &  \multicolumn{7}{c|}{Dataset samples} \\ \cline{5-11}
			&										   & 							 		   & 										  &		 
			\multicolumn{1}{c|}{Public} & \multicolumn{1}{c|}{Source} & \multicolumn{1}{c|}{Available} & \multicolumn{1}{c|}{Labeling} & \multicolumn{1}{c|}{Benign} & \multicolumn{1}{c|}{Malicious} & \multicolumn{1}{c|}{Total}\\ \hline\hline
			\endhead
			
			Bailey \textit{et al.}~\cite{BaileyOberheideAndersenEtAl2007} & Hierarchical clustering with normalized compression distance & APIs/System calls, file system, Windows Registry, and network  & Evasion techniques and samples requiring user interactions reduce the accuracy of the proposed classification method. The dataset is small. & \multicolumn{1}{c|}{$\xmark$}  & \multicolumn{1}{c|}{\makecell{Albor Malware\\Library and\\public\\repository}} & \multicolumn{1}{c|}{\xmark} & \multicolumn{1}{c|}{Automated}& \multicolumn{1}{c|}{$0$} & \multicolumn{1}{c|}{$8,228$} & \multicolumn{1}{c|}{$8,228$}\\ \hline
			Bayer \textit{et al.}~\cite{BayerComparettiHlauschekEtAl2009} & Clustering with locality sensitive hashing & APIs/System calls & Evasion techniques and samples requiring user interactions reduce the approach accuracy.& \multicolumn{1}{c|}{$\xmark$}  & \multicolumn{1}{c|}{\makecell{Anubis website}} & \multicolumn{1}{c|}{\xmark} & \multicolumn{1}{c|}{\makecell{Manual\\and\\automated}}& \multicolumn{1}{c|}{$0$} & \multicolumn{1}{c|}{$75,692$} & \multicolumn{1}{c|}{$75,692$}\\ \hline
			Rieck \textit{et al.}~\cite{RieckTriniusWillemsEtAl2011} & Prototype-based classification and clustering with Euclidean distance & Byte sequences and APIs/system calls &  Evasion techniques and samples requiring user interactions reduce the accuracy of the proposed framework.& \multicolumn{1}{c|}{$\xmark$}  & \multicolumn{1}{c|}{\makecell{CWSandbox\\and Sunbelt\\Software}} & \multicolumn{1}{c|}{\xmark} & \multicolumn{1}{c|}{Automated}& \multicolumn{1}{c|}{$0$} & \multicolumn{1}{c|}{$36,831$} & \multicolumn{1}{c|}{$36,831$}\\ \hline
			\multicolumn{11}{r}{\textit{Continue on the next page}} \\
			Palahan \textit{et al.}~\cite{PalahanBabicChaudhuriEtAl2013} & Logistic Regression & APIs/System calls & Evasion techniques and samples requiring user interactions reduce the accuracy of the proposed framework, while unknown observed behaviors are classified as malicious. The dataset used in the experimental evaluations is very small.& \multicolumn{1}{c|}{$\xmark$}  & \multicolumn{1}{c|}{\makecell{Own honeypot}} & \multicolumn{1}{c|}{\xmark} & \multicolumn{1}{c|}{$-$}& \multicolumn{1}{c|}{$49$} & \multicolumn{1}{c|}{$912$} & \multicolumn{1}{c|}{$961$}\\ \hline
			Egele \textit{et al.}~\cite{Egele:2014} & SVM\footnotemark & APIs/System calls, memory, and CPU registers & The accuracy of computed PE function similarities drops when different compiler toolchains or aggressive optimization levels are used. The dataset is small. & \multicolumn{1}{c|}{$\checkmark$}  & \multicolumn{1}{c|}{\makecell{\texttt{coreutils-8.13}\\program suite}} & \multicolumn{1}{c|}{\xmark} & \multicolumn{1}{c|}{$-$}& \multicolumn{1}{c|}{$1,140$} & \multicolumn{1}{c|}{$0$} & \multicolumn{1}{c|}{$1,140$}\\ \hline
		\end{longtable}
		
		\begin{longtable}{| m{2.5cm} | m{2.7cm} | m{2.5cm} | m{2.9cm} | m{.5} | m{.5cm} | m{.5cm} | m{.5cm} | m{.5cm} | m{.5cm} | m{.5cm}}
			\caption{Characterization of surveyed papers having malware differences detection as objective.}
			\label{tab:char_malware_differences_detection}
			\\ \hline
			\multirow{2}{*}{Paper} & \multirow{2}{*}{Algorithms} &  \multirow{2}{*}{Features} & \multirow{2}{*}{Limitations} &  \multicolumn{7}{c|}{Dataset samples} \\ \cline{5-11}
			&										   & 							 		   & 										  &		 
			\multicolumn{1}{c|}{Public} & \multicolumn{1}{c|}{Source} & \multicolumn{1}{c|}{Available} & \multicolumn{1}{c|}{Labeling} & \multicolumn{1}{c|}{Benign} & \multicolumn{1}{c|}{Malicious}  & \multicolumn{1}{c|}{Total}\\ \hline\hline
			\endfirsthead
			\caption{Characterization of surveyed papers having malware differences detection as objective. (Continued)} %
			\\ \hline
			\multirow{2}{*}{Paper} & \multirow{2}{*}{Algorithms} &  \multirow{2}{*}{Features} & \multirow{2}{*}{Limitations} &  \multicolumn{7}{c|}{Dataset samples} \\ \cline{5-11}
			&										   & 							 		   & 										  &		 
			\multicolumn{1}{c|}{Public} & \multicolumn{1}{c|}{Source} & \multicolumn{1}{c|}{Available} & \multicolumn{1}{c|}{Labeling} & \multicolumn{1}{c|}{Benign} & \multicolumn{1}{c|}{Malicious} & \multicolumn{1}{c|}{Total}\\ \hline\hline
			\endhead
			
			Bayer \textit{et al.}~\cite{BayerComparettiHlauschekEtAl2009} & Clustering with locality sensitive hashing & APIs/System calls & Evasion techniques and samples requiring user interactions reduce the approach accuracy.& \multicolumn{1}{c|}{$\xmark$}  & \multicolumn{1}{c|}{\makecell{Anubis website}} & \multicolumn{1}{c|}{\xmark} & \multicolumn{1}{c|}{\makecell{Manual\\and\\automated}}& \multicolumn{1}{c|}{$0$} & \multicolumn{1}{c|}{$75,692$} & \multicolumn{1}{c|}{$75,692$}\\ \hline
			Lindorfer \textit{et al.}~\cite{LindorferKolbitschComparetti2011} & Rule-based classifier & APIs/System calls and network & Sophisticated evasion techniques and samples requiring user interactions can still bypass detection processes. The dataset used in the experimental evaluations is small. & \multicolumn{1}{c|}{$\xmark$}  & \multicolumn{1}{c|}{\makecell{Anubis\\Sandbox}} & \multicolumn{1}{c|}{\xmark} & \multicolumn{1}{c|}{Automated} & \multicolumn{1}{c|}{$0$} & \multicolumn{1}{c|}{$1,871$} & \multicolumn{1}{c|}{$1,871$}\\ \hline
			Rieck \textit{et al.}~\cite{RieckTriniusWillemsEtAl2011} & Prototype-based classification and clustering with Euclidean distance & Byte sequences and APIs/system calls &  Evasion techniques and samples requiring user interactions reduce the accuracy of the proposed framework.& \multicolumn{1}{c|}{$\xmark$}  & \multicolumn{1}{c|}{\makecell{CWSandbox\\and Sunbelt\\Software}} & \multicolumn{1}{c|}{\xmark} & \multicolumn{1}{c|}{Automated}& \multicolumn{1}{c|}{$0$} & \multicolumn{1}{c|}{$36,831$} & \multicolumn{1}{c|}{$36,831$}\\ \hline
			\multicolumn{11}{r}{\textit{Continue on the next page}} \\
			Palahan \textit{et al.}~\cite{PalahanBabicChaudhuriEtAl2013} & Logistic Regression & APIs/System calls & Evasion techniques and samples requiring user interactions reduce the accuracy of the proposed framework, while unknown observed behaviors are classified as malicious. The dataset used in the experimental evaluations is small.& \multicolumn{1}{c|}{$\xmark$}  & \multicolumn{1}{c|}{\makecell{Own honeypot}} & \multicolumn{1}{c|}{\xmark} & \multicolumn{1}{c|}{$-$}& \multicolumn{1}{c|}{$49$} & \multicolumn{1}{c|}{$912$} & \multicolumn{1}{c|}{$961$}\\ \hline
			Santos \textit{et al.}~\cite{SantosBrezoUgarte-PedreroEtAl2013} & Decision Tree, \textit{k}-NN, Bayesian Network, Random Forest & Opcodes & Opcode sequence is not optimal and the dataset size is small. The proposed method is not effective against packed malware. The dataset is small. & \multicolumn{1}{c|}{$\checkmark$}  & \multicolumn{1}{c|}{\makecell{Own machines\\and\\VXHeavens}} & \multicolumn{1}{c|}{\xmark} & \multicolumn{1}{c|}{Automated}& \multicolumn{1}{c|}{$1,000$} & \multicolumn{1}{c|}{$1,000$} & \multicolumn{1}{c|}{$2,000$}\\ \hline
			\multicolumn{11}{r}{\textit{Continue on the next page}} \\
			Polino \textit{et al.}~\cite{Polino:2015} & Clustering with Jaccard similarity & APIs/System calls & Evasion techniques, packed malware, and samples requiring user interactions reduce the accuracy of the proposed framework. API calls sequence used to identify sample behaviors is not optimal. The dataset size is small. & \multicolumn{1}{c|}{$-$}  & \multicolumn{1}{c|}{\makecell{$-$}} & \multicolumn{1}{c|}{$-$} & \multicolumn{1}{c|}{$-$}& \multicolumn{1}{c|}{$?$} & \multicolumn{1}{c|}{$?$} & \multicolumn{1}{c|}{$2,136$}\\ \hline	
		\end{longtable}

	\scriptsize
	\centering
	\begin{longtable}{| m{2.5cm} | m{2.2cm} | m{2.5cm} | m{3cm} | m{1} | m{1.cm} | m{1cm} | m{1.2cm} | m{1cm} | m{1cm} | m{1cm}}
		\caption{Characterization of surveyed papers having malware category detection as objective. $^{4}$Instead of using machine learning techniques, these articles rely on Hidden Markov Models to detect metamorphic viruses~\cite{Wong2006,Attaluri2009}. } \\ \hline
		\label{tab:char_malware_category_detection}
		\multirow{2}{*}{Paper} & \multirow{2}{*}{Algorithms} &  \multirow{2}{*}{Features} & \multirow{2}{*}{Limitations} &  \multicolumn{7}{c|}{Dataset samples} \\ \cline{5-11}
		&										   & 							 		   & 										  &		 
		\multicolumn{1}{c|}{Public} & \multicolumn{1}{c|}{Source} & \multicolumn{1}{c|}{Available} & \multicolumn{1}{c|}{Labeling} & \multicolumn{1}{c|}{Benign} & \multicolumn{1}{c|}{Malicious}  & \multicolumn{1}{c|}{Total}\\ \hline\hline
		\endfirsthead
		\caption{Characterization of surveyed papers having malware category detection as objective. (Continued)}                              
		\\ \hline
		\multirow{2}{*}{Paper} & \multirow{2}{*}{Algorithms} &  \multirow{2}{*}{Features} & \multirow{2}{*}{Limitations} &  \multicolumn{7}{c|}{Dataset samples} \\ \cline{5-11}
		&										   & 							 		   & 										  &		 
		\multicolumn{1}{c|}{Public} & \multicolumn{1}{c|}{Source} & \multicolumn{1}{c|}{Available} & \multicolumn{1}{c|}{Labeling} & \multicolumn{1}{c|}{Benign} & \multicolumn{1}{c|}{Malicious}  & \multicolumn{1}{c|}{Total}\\ \hline\hline
		\endhead
		Wong and \newline Stamp~\cite{Wong2006} & -\footnotemark & Opcodes & Detection fails if metamorphic malware are similar to benign files. The dataset is extremely small. & \multicolumn{1}{c|}{$\xmark$}  & \multicolumn{1}{c|}{\makecell{Cygwin\\and\\VX Heavens\\generators}} & \multicolumn{1}{c|}{\xmark} & \multicolumn{1}{c|}{$-$}& \multicolumn{1}{c|}{$40$} & \multicolumn{1}{c|}{$25$} & \multicolumn{1}{c|}{$65$}\\ \hline
		Attaluri \textit{et al.}~\cite{Attaluri2009} & -$^6$ & Opcodes & The proposed approach is not effective against all types of metamorphic viruses. The dataset size is very small. & \multicolumn{1}{c|}{$\xmark$}  & \multicolumn{1}{c|}{\makecell{Cygwin,\\legitimate\\DLLs and\\VX Heavens\\generators}} & \multicolumn{1}{c|}{\xmark} & \multicolumn{1}{c|}{$-$}& \multicolumn{1}{c|}{$240$} & \multicolumn{1}{c|}{$70$} & \multicolumn{1}{c|}{$310$}\\ \hline
		Tian \textit{et al.}~\cite{Tian:2008} & Rule-based classifier & Function length & Function lengths alone are not sufficient to detect Trojans and the dataset used in the experimental evaluations is very small.& \multicolumn{1}{c|}{$-$}  & \multicolumn{1}{c|}{\makecell{$-$}} & \multicolumn{1}{c|}{$-$} & \multicolumn{1}{c|}{$-$}& \multicolumn{1}{c|}{$0$} & \multicolumn{1}{c|}{$721$} & \multicolumn{1}{c|}{$721$}\\ \hline
		Siddiqui \textit{et al.}~\cite{Siddiqui:2009} & Decision Tree, Random Forest & Opcodes & Advanced packing techniques could reduce detection accuracy. The dataset used in the experimental evaluations is small. & \multicolumn{1}{c|}{$\checkmark$}  & \multicolumn{1}{c|}{\makecell{Windows XP\\and\\VX Heavens}} & \multicolumn{1}{c|}{\xmark} & \multicolumn{1}{c|}{$-$} & \multicolumn{1}{c|}{$1,444$} & \multicolumn{1}{c|}{$1,330$} & \multicolumn{1}{c|}{$2,774$}\\ \hline
		Chen \textit{et al.}~\cite{ChenRoussopoulosLiangEtAl2012} & Random Forest, SVM & Byte sequences & The proposed framework heavily relies on security companies' encyclopedias.& \multicolumn{1}{c|}{$\xmark$}  & \multicolumn{1}{c|}{\makecell{Trend Micro}} & \multicolumn{1}{c|}{\xmark} & \multicolumn{1}{c|}{$-$} & \multicolumn{1}{c|}{$0$} & \multicolumn{1}{c|}{$330,248$} & \multicolumn{1}{c|}{$330,248$}\\ \hline
		\multicolumn{11}{r}{\textit{Continue on the next page}} \\ 
		Comar \textit{et al.}~\cite{Comar:2013} & Random Forest, SVM & Network & Network features are extracted by a commercial traffic analyzer. & \multicolumn{1}{c|}{$\xmark$}  & \multicolumn{1}{c|}{\makecell{Internet Service\\Provider}} & \multicolumn{1}{c|}{\xmark} & \multicolumn{1}{c|}{\makecell{Manual and\\automated}} & \multicolumn{1}{c|}{$212,505$} & \multicolumn{1}{c|}{$4,394$} & \multicolumn{1}{c|}{$216,899$}\\ \hline
		Kwon \textit{et al.}~\cite{Kwon2015} & Random Forest & Network & Not able to detect bots with rootkit capabilities. & \multicolumn{1}{c|}{$\checkmark$}  & \multicolumn{1}{c|}{\makecell{Symantec’s\\Worldwide\\ Intelligence\\Network \\Environment}} & \multicolumn{1}{c|}{\xmark} & \multicolumn{1}{c|}{$-$}  & \multicolumn{1}{c|}{$?$} & \multicolumn{1}{c|}{$?$} & \multicolumn{1}{c|}{$24 * 10^6$}\\ \hline
		Sexton \textit{et al.}~\cite{Sexton2015} & Rule-based classifier, Logistic Regression, \newline Na{\"i}ve Bayes, SVM & Byte sequences and opcodes & Obfuscation techniques reduce detection accuracy. The dataset used in the experimental evaluations is small. & \multicolumn{1}{c|}{$-$}  & \multicolumn{1}{c|}{$-$} & \multicolumn{1}{c|}{$-$} & \multicolumn{1}{c|}{$-$}& \multicolumn{1}{c|}{$4,622$} & \multicolumn{1}{c|}{$197$} & \multicolumn{1}{c|}{$4,819$}\\ \hline
	\end{longtable}
\end{landscape}

\section{Issues and Challenges} \label{sec:issues_challenges}

Based on the characterization detailed in section~\ref{sec:characterization}, this section identifies the main issues and challenges of surveyed papers. In the specific, the main problems regard the usage of anti-analysis techniques by malware (\S~\ref{sec:ic_anti_analysis}), what operation set to consider~\ref{sec:ic_operation_set} and used dataset~\ref{sec:ic_datasets}.

\subsection{Anti-analysis Techniques} \label{sec:ic_anti_analysis}
Malware developers want to avoid their samples to be analysed, so they devise and refine several \textit{anti-analysis techniques} that are effective in hindering the reverse engineering of executables. Indeed many surveyed works claim that the solution they propose does not work or loses in accuracy when samples using such techniques are considered (\S~\ref{sec:characterization}).

Static analysis (\S~\ref{sec:taxonomy_feature_extraction}) is commonly prevented by rendering sample binary and resources unreadable through \textit{obfuscation}, \textit{packing} or \textit{encryption}. Anyway, at runtime, code and any other concealed data has to be either deobfuscated, unpacked or decrypted to enable the correct execution of the payload. This implies that such a kind of anti-analysis techniques can be overcome by using dynamic analysis (\S~\ref{sec:taxonomy_feature_extraction}) to make the sample unveil hidden information and load them in memory, where they can then be extracted by creating a dump. Refer to Ye \textit{et al.}~\cite{Ye2017} for a detailed disquisition on how obfuscation, packing and encryption are used by malware developers.

More advanced anti-analysis techniques exist to keep malware internals secret even when dynamic analysis is used. One approach, commonly referred to as \textit{environmental awareness}, consists in the malware trying to detect whether it is being executed in a controlled setting where an analyst is trying to dissect it, for example by using a virtual machine or by running the sample in debug mode. %
If any cue is found of possibly being under analysis, then the malware does not execute its malicious payload.
Miramirkhani \textit{et al.}~\cite{Miramirkhani2017} show that a malware can easily understand if it is running into an artificial environment.
Other approaches rely on \textit{timing-based evasion}, i.e. they only show their malicious behaviour at predetermined dates and times. Other malware instead require or wait for some \textit{user interaction} to start their intended activity, in order to make any kind of automatic analysis infeasible.

Identifying and overcoming these anti-analysis techniques is an important direction to investigate to improve the effectiveness of malware analysis.
Recent academic and not-academic literature are aligned on this aspect. Karpin and Dorfman~\cite{Karpin2017} highlight the need to address very current problems such as discovering where malware configuration files are stored and whether standard or custom obfuscation/packing/encryption algorithms are employed. Deobfuscation~\cite{Blazytko2017,Kotov2018} and other operations aimed at supporting binary reverse engineering, such as function similarity identification~\cite{Liao2018}, are still very active research directions. Symbolic execution techniques~\cite{SurveySymExec-CSUR18} are promising means to understand what execution paths trigger the launch of the malicious payload.

\subsection{Operation Set} \label{sec:ic_operation_set}

Opcodes, instructions, APIs and system calls (hereinafter, we refer to them in general as \textit{operations}) are the most used and powerful features employed for malware analysis (\S~\ref{sec:characterization}), as they allow to directly and accurately model sample behaviour. Normally, to reduce complexity and required computational power, only a subset of all the available operations is considered. %
This choice of ignoring part of the operations at disposal can reduce the accuracy of malware behaviour model, which in turn reflects on the reliability of analysis outcomes. This issue has been raised explicitly in some surveyed papers, including~\cite{AndersonStorlieLane2012, Gharacheh:2015, KhodamoradiFazlaliMardukhiEtAl2015}, while others are anyway affected although they do not mention it, such as~\cite{Ghiasi:2015, LiangPangDai2016, HuangYeJiang2009}.

On one hand, this challenge can be addressed by either improving or using different machine learning techniques to achieve a more effective feature selection. On the other hand, program analysis advances can be leveraged to enhance the accuracy of disassemblers and decompilers, indeed these tools are known to be error-prone~\cite{Guilfanov2008, Rosseau2018} and are thus likely to affect negatively the whole analyses. Approaches that improve the quality of generated disassembly and decompiled code, as in~\cite{eschulte2018}, can reduce the impact due to these errors.

\subsection{Datasets} \label{sec:ic_datasets}

More than $72$\% of surveyed works use datasets with both malicious and benign samples, while about $28$\% rely on datasets with malware only.
Just two works rely on benign datasets only~\cite{Egele:2014,Caliskan-Islam:2015}, because their objectives are identifying sample similarities and attributing the ownership of some source codes under analysis, respectively. %

\begin{figure}[b!]
	\centering
	\includegraphics[width=\linewidth]{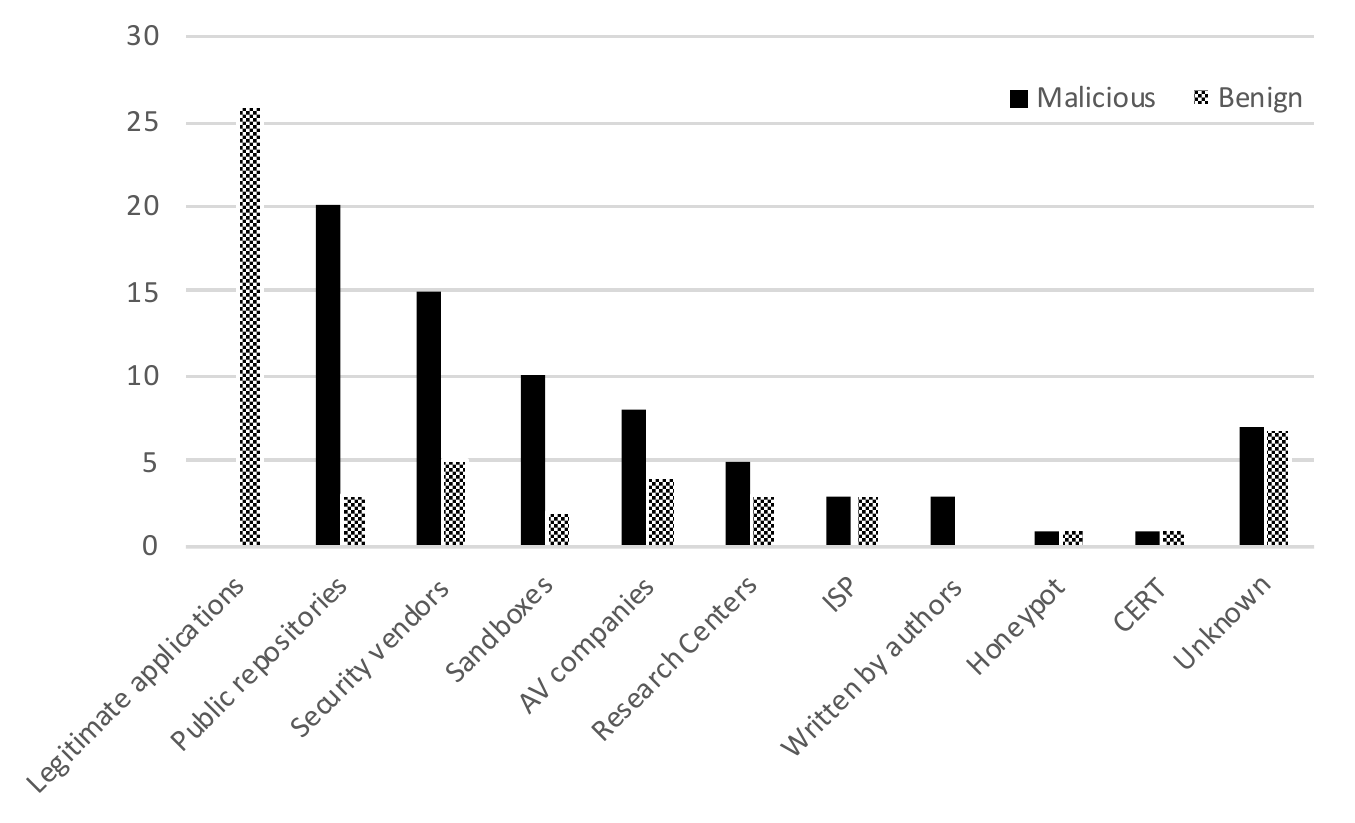}
	\captionof{figure}{Frequency histogram showing how many reviewed papers use each type of source (e.g. public repositories, honeypot) to collect their datasets, and whether it is used to gather malware or benign samples.}
	\label{fig:mal_ben_datasets}
\end{figure}

Figure~\ref{fig:mal_ben_datasets} shows the dataset sources for malicious and benign samples. %
It is worth noting that most of benign datasets consists of legitimate applications (e.g. software contained in ``Program Files'' or ``system'' folders), while most of malware have been obtained from public repositories, security vendors and popular sandboxed analysis services.
The most popular public repository in the examined works is VX Heavens~\cite{vxheaven}, followed by Offensive Computing~\cite{offensive_computing} and Malicia Project~\cite{malicia_project}.
The first two repositories are still actively maintained at the time of writing, while Malicia Project has been permanently shut down due to dataset ageing and lack of maintainers.

Security vendors, popular sandboxed analysis services, and AV companies have access to a huge number of samples.
Surveyed works rely on CWSandbox, developed by ThreatTrack Security~\cite{threattrack}, and Anubis~\cite{anubis}. %
As can be observed from Figure~\ref{fig:mal_ben_datasets}, these sandboxes are mainly used for obtaining malicious samples.
Internet Service Providers (ISPs), honeypots and Computer Emergency Response Teams (CERTs) share with researchers both benign and malicious datasets.
A few works use malware developed by the authors~\cite{Gharacheh:2015,KhodamoradiFazlaliMardukhiEtAl2015}, created using malware toolkits~\cite{Wong2006} such as Next Generation Virus Constrution Kit, Virus Creation Lab, Mass Code Generator and Second Generation Virus Generator, all available on VX Heavens~\cite{vxheaven}.
A minority of analysed papers do not mention the source of their datasets.

Among surveyed papers, a recurring issue is the size of used dataset. Many works, including~\cite{Kolter:2006, Ahmed2009, FirdausiLimErwinEtAl2010}, carry out evaluations on less than $1,000$ samples. 
Just $39\%$ of reviewed studies test their approaches on a population greater than $10,000$ samples.

When both malicious and benign samples are used for the evaluation, it is crucial to reflect their real distribution~\cite{Shultz:2001,Kolter:2006,Ahmed2009,Siddiqui:2009,FirdausiLimErwinEtAl2010,ParkReevesMulukutlaEtAl2010,AndersonQuistNeilEtAl2011,Santos:2011,AndersonStorlieLane2012,BilgeBalzarottiRobertsonEtAl2012,Yonts2012,DahlStokesDengEtAl2013,EskandariKhorshidpourHashemi2013,IslamTianBattenEtAl2013,Kirat2013,PalahanBabicChaudhuriEtAl2013,SantosBrezoUgarte-PedreroEtAl2013,SantosDevesaBrezoEtAl2013,BaiWangZou2014,UppalSinhaMehraEtAl2014,ElhadiMaarofBarry2015,FengXiongCaoEtAl2015,Ghiasi:2015,KawaguchiOmote2015,Lin:2015,Mao2015,Pai:2015,Saxe2015,SrakaewPiyanuntcharatsr:2015,WuechnerOchoaPretschner2015,Raff2017}. Indeed, there needs to be a huge imbalance because non-malware executables are the overwhelming majority.
48\% of surveyed works do not take care of this aspect and use datasets that either are balanced between malware and non malicious software or, even, have more of the former than the latter.
In~\cite{Yonts2012}, Yonts supports his choice of using a smaller benign dataset by pointing out that changes in standard system files and legitimate applications are little.
$38\%$ of examined papers employ instead datasets having a proper distribution of malware and non-malware, indeed they are either unbalanced towards benign samples or use exclusively benign or malicious software. As an example, the majority of surveyed papers having malware similarities detection as objective (see Table~\ref{tab:char_malware_similarities_detection}) contains exclusively either malware or legitimate applications~\cite{BaileyOberheideAndersenEtAl2007,BayerComparettiHlauschekEtAl2009,RieckTriniusWillemsEtAl2011,Egele:2014}.
The remaining $14\%$ does not describe how datasets are composed.

Differently from other research fields, no reference benchmark is available for malware analysis to compare accuracy and performance with other works. Furthermore, published results are known to be biased towards good results~\cite{Sanders2017}.
In addition, since the datasets used for evaluations are rarely shared, it is nearly impossible to compare works.
Only two surveyed works have shared their dataset~\cite{Shultz:2001,UpchurchZhou2015}, while a third one plans to share it in the future~\cite{MohaisenAlrawiMohaisen2015}. It is worth mentioning that one of the shared dataset is from 2001, hence almost useless today.
Indeed, temporal information is crucial to evaluate malware analysis results~\cite{Miller2015} and determine whether machine learning models have become obsolete~\cite{Jordaney2017,Harang2018}.

Given such lack of reference datasets, we propose three desiderata for malware analysis benchmarks.
\begin{enumerate}
	
	\item Benchmarks should be labeled accordingly to the specific objectives to achieve.
	As an example, benchmarks for families selection should be labeled with samples' families. %
	\item Benchmarks should model realistically the sample distributions of real-world scenarios, considering the objectives to attain.
	For example, benchmarks for malware detection should contain a set of legitimate applications orders of magnitude greater than the number of malware samples. %
	\item Benchmarks should be actively maintained and updated over time with new samples, trying to keep pace with the malware industry.
	Samples should also be provided with temporal information, e.g., when they have been spotted first.
\end{enumerate}
Datasets used in~\cite{Shultz:2001} and~\cite{UpchurchZhou2015} are correctly labeled according to malware detection and malware variants selection objectives, respectively.
Both datasets are not balanced.
In Shultz \textit{et al.}~\cite{Shultz:2001}, the described dataset is biased towards malicious programs, while in~\cite{UpchurchZhou2015} diverse groups of variants contain a different number of samples, ranging from $3$ to $20$.
Finally, analysed datasets are not actively maintained and do not contain any temporal information (in~\cite{Shultz:2001}, the authors do not mention if such information has been included into the dataset).

\section{Topical Trends} \label{sec:topical_trends}

This section outlines a list of topical trends in malware analysis, i.e. topics that are currently being investigated but have not reached the same level of maturity of the other areas described in previous sections.

\subsection{Malware Development Detection} \label{sec:topical_trends_development_detection}

Malware developers can use online public services like VirusTotal~\cite{virtustotal} and Malwr~\cite{malwr} to test the effectiveness of their samples in evading most common antiviruses.
Malware analysts can leverage such behaviour by querying these online services to obtain additional information useful for the analysis, such as submission time and how many online antiviruses classify a sample as malicious. %
Graziano \textit{et al.}~\cite{Graziano:2015} leverage submissions to an online sandbox for identifying cases where new samples are being tested, with the final aim to detect novel malware during their development process.
Surprisingly, it turned out that samples used in infamous targeted campaigns had been submitted to public sandboxes months or years before.

With reference to the proposed taxonomy, advances in the state of the art in malware analysis could be obtained by analysing submissions to online malware analysis services, to extract additional machine learning features and gather intelligence on what next malware are likely to be.

\subsection{Malware Attribution} \label{sec:topical_trends_objectives_attribution}

Another aspect of interest for malware analysts is the identification of who developed a given sample, i.e. the attribution of a malware to a specific malicious actor.
There are a number of features in a binary to support this process: 
used programming language, included IP addresses and URLs, and the language of comments and resources.
Additional, recently proposed features which can be used for attribution are the time slot when the malware communicates with a command and control centre and what digital certificates are used~\cite{Ruthven2017}.
Features related to the coding style can also reveal important details on developer's identity, at least for arguing whether different malware have been developed by the same person or group. In~\cite{Caliskan-Islam:2015}, the author's coding style of a generic software (i.e. not necessarily malicious) is accurately profiled through syntactic, lexical, and layout features.
Unfortunately, this approach requires the availability of source code, which happens only occasionally, e.g. in case of leaks and/or public disclosures. %

Malware attribution can be seen as an additional analysis objective, according to the proposed taxonomy. Progresses in this direction through machine learning techniques are currently hindered by the lack of ground truth on malware authors, which proves to be really hard to provision. Recent approaches leverage on public reports referring to APT groups and detailing what malware they are supposed to have developed: those reports are parsed to mine the relationships between malicious samples and corresponding APT group authors~\cite{10.1007/978-3-319-60080-2_21}. The state of the art in malware attribution through machine learning can be advanced by researching alternative methods to generate reliable ground truth on malware developers, or on what malware have been developed by the same actor.

\subsection{Malware Triage} \label{sec:topical_trends_objectives_triage}

Given the huge amount of new malware that need to be analysed, a fast and accurate prioritisation is required to identify what samples deserve more in depth analyses. %
This can be decided on the basis of the level of similarity with already known samples. If a new malware resembles very closely other binaries that have been analysed before, then its examination is not a priority. Otherwise, further analyses can be advised if a new malware really looks differently from everything else observed so far.
This process is referred to as \textit{malware triage} and shares some aspects with malware similarity analysis, as they both provide key information to support malware analysis prioritisation. Anyway, they are different because triage requires faster results at the cost of worse accuracy, hence different techniques are usually employed~\cite{JangBrumleyVenkataraman2011,Kirat2013,10.1007/978-3-319-60080-2_21,Rosseau2018}. %

Likewise attribution, triage can be considered as another malware analysis objective. One important challenge of malware triage is finding the proper trade-off between accuracy and performance, which fits with the problems we address in the context of malware analysis economics (see Section~\ref{sec:malware_analysis_economics}).

\subsection{Prediction of Future Variants} \label{sec:topical_trends_prediction_future_variants}

Compared to malware analysts, malware developers have the advantage of knowing current anti-malware measures and thus novel variants can be designed accordingly. A novel trend in malware analysis is investigating the feasibility to fill that gap by predicting how future malware will look like, so as to allow analysts to update anti-malware measures ahead. Howard \textit{et al.}~\cite{8323965} use machine learning techniques to model patterns in malware family evolutions and predict future variants.

This problem can be seen as yet another objective in the malware analysis taxonomy. It has not been investigated much yet, only a couple of works seem to address that topic~\cite{juzonis2012specialized, 8323965}. Given its great potential to proactively identify novel malware, and considering the opportunity to exploit existing malware families datasets, we claim the worthiness to boost the research on malware evolution prediction through machine learning techniques.

\subsection{Other Features} \label{sec:topical_trends_features}

This section describes features different from those analysed in Section~\ref{sec:taxonomy_pe_feature} and that have been used by just a few papers so far. In view of advancing the state of the art in malware analysis, additional research is required on the effectiveness of using such features to improve the accuracy of machine learning techniques.

\subsubsection{Memory Accesses} \label{sec:topical_trends_features_memory_accesses}

Any data of interest, such as user generated content, %
is temporary stored in main memory, hence analysing how memory is accessed can reveal important information about the behaviour of an executable~\cite{Hal:2012}. %
Kong~\textit{et al.} rely on statically trace reads and writes in main memory~\cite{Kong:2013}, while 
Egele \textit{et al.} dynamically trace values read from and written to stack and heap~\cite{Egele:2014}.

\subsubsection{Function Length} \label{sec:topical_trends_features_function_length}

Another characterising feature is the function length, measured as the number of bytes contained in a function.
This input alone is not sufficient to discriminate malicious executables from benign software, indeed it is usually combined with other features. %
This idea, formulated in~\cite{Tian:2008}, is adopted in~\cite{IslamTianBattenEtAl2013}, where function length frequencies, extracted through static analysis, are used together with other static and dynamic features.

\subsubsection{Raised Exceptions} \label{sec:topical_trends_features_raised_exceptions}

The analysis of the exceptions raised during the execution can help understanding what strategies a malware adopts to evade analysis systems~\cite{SantosDevesaBrezoEtAl2013,Asquith2015}.
A common trick to deceive analysts is throwing an exception to run a malicious handler, registered at the beginning of malware execution.
In this way, examining the control flow becomes much more complex.

\section{Malware Analysis Economics} \label{sec:malware_analysis_economics}

Analysing samples through machine learning techniques requires complex computations for extracting desired features and running chosen algorithms. 
The time complexity of these computations has to be carefully taken into account to ensure they complete fast enough to keep pace with the speed new malware are developed. 
Space complexity has to be considered as well, indeed feature space can easily become excessively large (e.g., using n-grams), and also the memory required by machine learning algorithms can grow to the point of saturating available resources.

Time and space complexities can be either reduced to adapt to processing and storage capacity at disposal, or they can be accommodated by supplying more resources. In the former case, the analysis accuracy is likely to worsen, while, in the latter, accuracy levels can be preserved at the cost of providing more computing machines, storage and network. There exist therefore trade-offs between maintaining high accuracy and performance of malware analysis on one hand, and supplying the required equipment on the other.
We refer to the study of these trade-offs as \textit{malware analysis economics}, and in this section we provide some initial qualitative discussions on this novel topic.\\

The time needed to analyse a sample through machine learning is mainly spent in feature extraction and algorithm execution. While time complexity of machine learning algorithms is widely discussed in literature, the same does not apply for the study of feature extraction execution time. The main aspect to take into account is whether desired features come from static or dynamic analysis, which considerably affects execution time because the former does not require to run the samples, while the latter does. %

\begin{table*}[b]
	\centering
	\tiny
	\caption{Type of analysis required for extracting the inputs presented in Sections~\ref{sec:taxonomy_pe_feature} and~\ref{sec:topical_trends_features}: strings, byte sequences, opcodes, APIs/system calls, file system, CPU registers, PE file characteristics, network, AV/Sandbox submissions, code stylometry, memory accesses, function length, and raised exceptions.}
	\label{tab:inputs_analysis}
	\begin{tabular}{| M{1cm} | M{0.65cm} | M{0.65cm} | M{0.65cm} | M{0.65cm} | M{0.65cm} | M{0.65cm} | M{0.65cm} | M{0.65cm} | M{0.65cm} | M{0.65cm} | M{0.65cm} | M{0.65cm} | M{0.65cm} | M{0.65cm} |}
		\hline
		Analysis & Str. & Byte \newline seq. & Ops & APIs \newline Sys. calls & File \newline sys. & CPU \newline reg.  & PE file \newline char.  & Net. &  Sub-mis. & Code \newline stylo. & Mem. &  Func. \newline len. &  Exc. \\ \hline\hline
		
		Static      & $\checkmark$ & $\checkmark$ & $\checkmark$ & $\checkmark$ & & $\checkmark$  & $\checkmark$ & & $\checkmark$ & $\checkmark$ & & $\checkmark$ & 
		\\\hline
		Dynamic & & & & $\checkmark$ & $\checkmark$ & $\checkmark$ &  & $\checkmark$  & $\checkmark$ & & $\checkmark$ &  &
		$\checkmark$  \\\hline
	\end{tabular}
\end{table*}

To deepen even further this point, Table~\ref{tab:inputs_analysis} reports for each feature type whether it can be extracted through static or dynamic analysis.
It is interesting to note that certain types of features can be extracted both statically and dynamically, with significant differences on execution time as well as on malware analysis accuracy. Indeed, while certainly more time-consuming, dynamic analysis enables to gather features that contribute relevantly to the overall analysis effectiveness~\cite{damodaran2015comparison}.
As an example, we can consider the features derived from API calls (see Table~\ref{tab:inputs_analysis}), which can be obtained both statically and dynamically.
Tools like IDA provide the list of imports used by a sample and can statically trace what API calls are present in the sample code.
Malware authors usually hide their suspicious API calls by inserting in the source code a huge number of legitimate APIs.
By means of dynamic analysis, it is possible to obtain the list of the APIs that the sample has actually invoked, thus simplifying the identification of those suspicious APIs.
By consequences, in this case dynamic analysis is likely to generate more valuable features compared to static analysis.
MazeWalker~\cite{Kulakov2017} is a typical example of how dynamic information can integrate static analysis. %

Although choosing dynamic analysis over, or in addition to, static seems obvious, its inherently higher time complexity constitutes a potential performance bottleneck for the whole malware analysis process, which can undermine the possibility to keep pace with malware evolution speed. The natural solution is to provision more computational resources to parallelise analysis tasks and thus remove bottlenecks. In turn, such solution has a cost to be taken into account when designing a malware analysis environment, such as the one presented by Laurenza \textit{et al.}~\cite{Laurenza2016}.

The qualitative trade-offs we have identified are between accuracy and time complexity (i.e., higher accuracy requires larger times), between time complexity and analysis pace (i.e., larger times implies slower pace), between analysis pace and computational resources (faster analysis demands using more resources), and between computational resources and economic cost (obviously, additional equipment has a cost). 
Similar trade-offs also hold for space complexity. %
As an example, when using n-grams as features, it has been shown that larger values of $n$ lead to more accurate analysis, at cost of having the feature space grow exponentially with $n$~\cite{UppalSinhaMehraEtAl2014,Lin:2015}. As another example, using larger datasets in general enables more accurate machine learning models and thus better accuracy, provided the availability of enough space to store all the samples of the dataset and the related analysis reports.

\begin{table*}[b]
	\centering
	\tiny
	\caption{Relationship between $n$ and number of features.}
	\label{tab:n_features}
	\begin{tabular}{| c | c |}
		\hline
		$n$ & feature count \\\hline\hline
		1 & 187 \\\hline
		2 & 6740 \\\hline
		3 & 46216 \\\hline
		4 & 130671 \\\hline
		5 & 342663 \\\hline
	\end{tabular}
\end{table*}

We present a qualitative, simplified example of analysis that leverages on the trade-offs just introduced. The scenario we target regards detecting malware families of new malicious samples (\S~\ref{sec:taxonomy_objective_malware_similarity_analysis}) using as features n-grams computed over invoked APIs (\S~\ref{sec:taxonomy_pe_feature}), recorded through dynamic analysis (\S~\ref{sec:taxonomy_feature_extraction}). We want here to explore the trade-offs between family detection accuracy, execution time, analysis pace and cost, in terms of required computational resources. For what concerns the scenario and qualitative numbers on the relationships between $n$, the number of features, accuracy and execution time, we take inspiration from the experimental evaluation presented by Lin \textit{et al.}~\cite{Lin:2015}. Table~\ref{tab:n_features} shows the relationship between $n$ and feature count. We introduce a few simplifying assumptions and constraints to make this qualitative example as consistent as possible. We assume that the algorithm used to detect families is parallelisable and ideally scalable, meaning that by doubling available machines we also double the throughput,  
i.e. the number of malware analysed per second. We want to process one million malware per day with an accuracy of at least $86\%$.
\begin{figure}
	\centering
	\includegraphics[width=\linewidth]{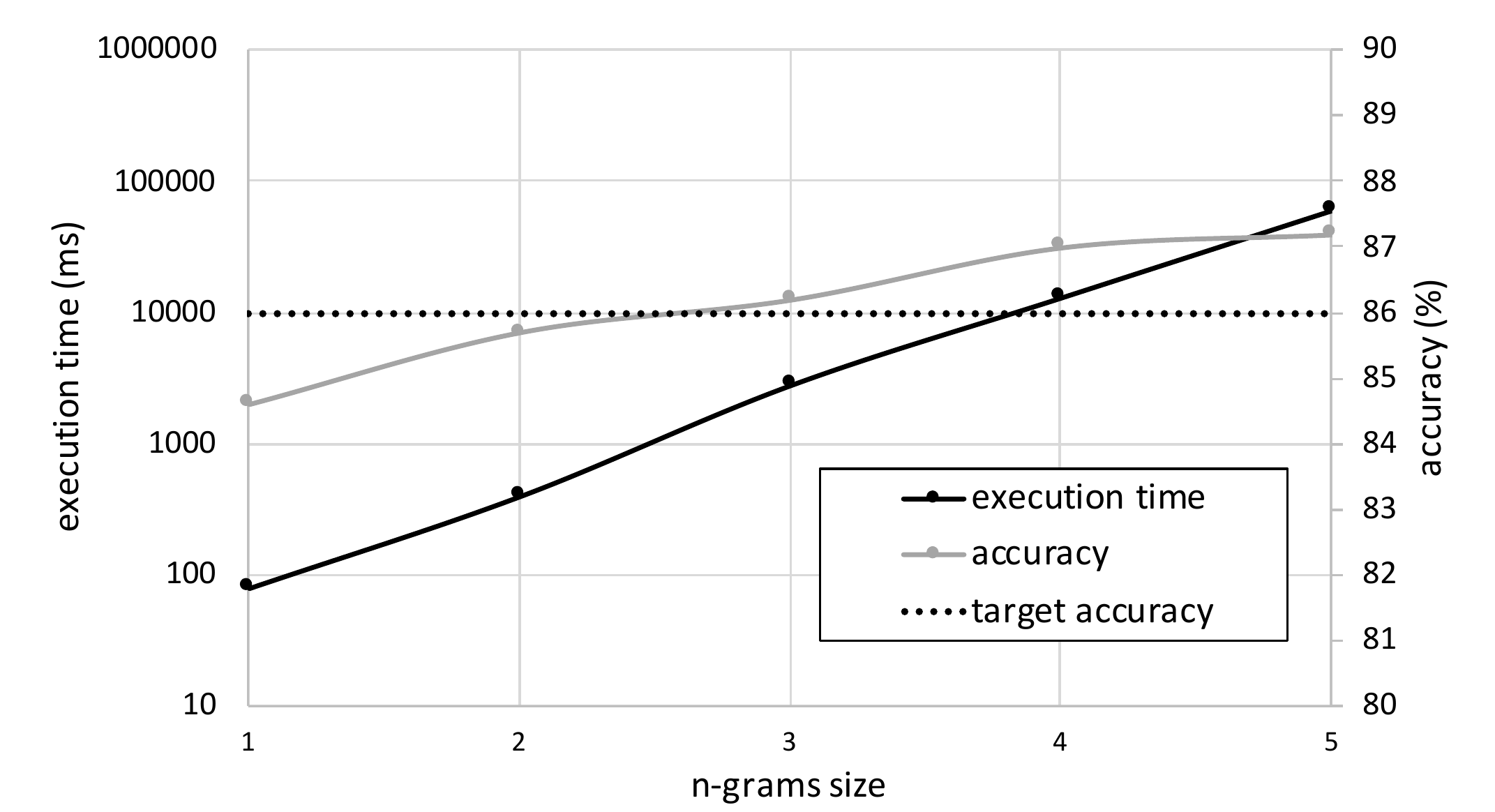}
	\captionof{figure}{Relationship between execution time (in logarithmic scale) and detection accuracy as $n$ varies. The target accuracy of $86\%$ is also reported.}
	\label{fig:time_accuracy}
\end{figure}

\begin{figure}
	\centering
	\includegraphics[width=\linewidth]{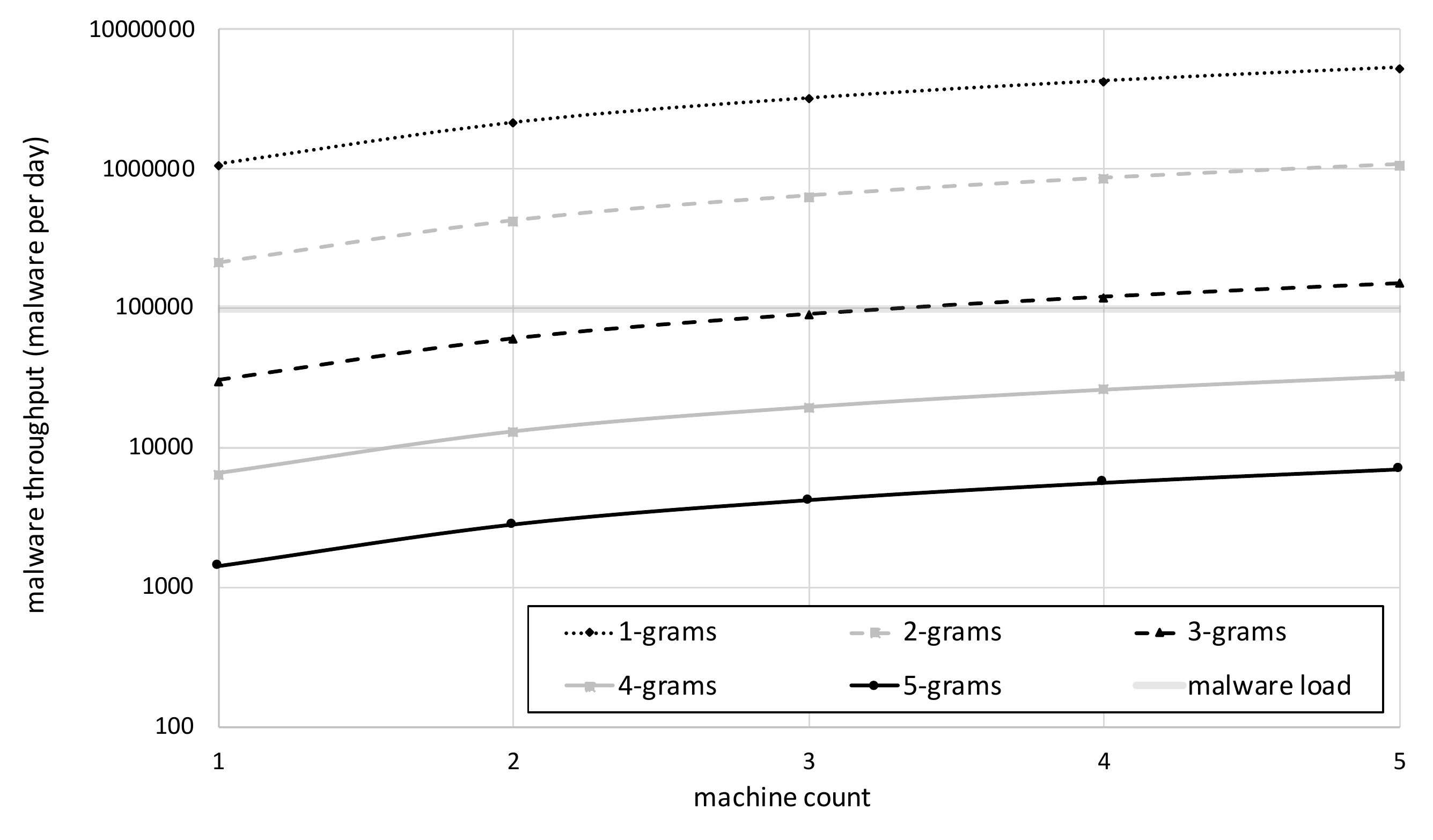}
	\captionof{figure}{Relationship between machine count and malware throughput (in logarithmic scale) for different n-grams sizes. The one million malware per day to sustain is also reported.}
	\label{fig:throughput}
\end{figure}

Figure~\ref{fig:time_accuracy} highlights the trade-off between execution time (in logarithmic scale) and detection accuracy as $n$ is varied. As $n$ grows, the accuracy increases almost linearly while the execution time has an exponential rise, which translates to an exponential decrease of how many malware per second can be processed. It can be noted that the minimum n-grams size to meet the accuracy requirement of 86\% is 3. The trade-off between analysis pace and cost can be observed in Figure~\ref{fig:throughput} where, by leveraging on the assumption of ideal scalability of the detection algorithm, it is shown that sustainable malware throughput (in logarithmic scale) increases linearly as the algorithm is parallelised on more machines. 4-grams and 5-grams cannot be used to cope with the expected malware load of one million per day, at least when considering up to five machines. On the other hand, by using four machines and 3-grams, we can sustain the target load and at the same time meet the constraint on detection accuracy.

The presented toy example is just meant to better explain how malware analysis economics can be used in practical scenarios. We claim the significance of investigating these trade-offs more in detail, with the aim of outlining proper guidelines and strategies to design a malware analysis environment in compliance with requirements on analysis accuracy and pace, while respecting budget constraints.

\section{Conclusion} \label{sec:conclusion}

We presented a survey on existing literature on malware analysis through machine learning techniques. There are five main contributions of our work. First, we proposed an organization of reviewed works according to three orthogonal dimensions: \textit{the objective of the analysis}, \textit{the type of features extracted from samples}, the \textit{machine learning algorithms used to process these features}.
Such characterization provides an overview on how machine learning algorithms can be employed in malware analysis, emphasising which specific feature classes allow to achieve the objective(s) of interest.
Second, we have arranged existing literature on PE malware analysis through machine learning according the proposed taxonomy, providing a detailed comparative analysis of surveyed works.
Third, we highlighted the current issues of machine learning for malware analysis: anti-analysis techniques used by malware, what operation set to consider for the features and used datasets.
Fourth, we identified topical trends on interesting objectives and features, such as malware attribution and triage.
Fifth, we introduced the novel concept of \textit{malware analysis economics}, concerning the investigation and exploitation of existing trade-offs between performance metrics of malware analysis (e.g., analysis accuracy and execution time) and economical costs.

Noteworthy research directions to investigate can be linked to those contributions.
Novel combinations of objectives, features and algorithms can be investigated to achieve better accuracy compared to the state of the art.
Moreover, observing that some classes of algorithms have never been used for a certain objective may suggest novel directions to examine further.
The discussion on malware analysis issues can provide further ideas worth to be explored. In particular, defining appropriate benchmarks for malware analysis is a priority of the whole research area.
The novel concept of malware analysis economics can encourage further research directions, where appropriate tuning strategies can be provided to balance competing metrics (e.g. accuracy and cost) when designing a malware analysis environment.

\section*{Acknowledgment}
This work has been partially supported by a grant of the Italian Presidency of Ministry Council and by the Laboratorio Nazionale of Cyber Security of the CINI (Consorzio Interuniversitario Nazionale Informatica).

\bibliography{references}

\begin{thebibliography}{100}
\expandafter\ifx\csname url\endcsname\relax
  \def\url#1{\texttt{#1}}\fi
\expandafter\ifx\csname urlprefix\endcsname\relax\def\urlprefix{URL }\fi
\expandafter\ifx\csname href\endcsname\relax
  \def\href#1#2{#2} \def\path#1{#1}\fi

\bibitem{Ye2017}
Y.~Ye, T.~Li, D.~Adjeroh, S.~S. Iyengar, A survey on malware detection using
  data mining techniques, ACM Computing Surveys (CSUR) 50~(3) (2017) 41.

\bibitem{av-test-report-2017}
{AV-TEST}, {Security Report 2016/17},
  \url{https://www.av-test.org/fileadmin/pdf/security_report/AV-TEST_Security_Report_2016-2017.pdf}
  (2017).

\bibitem{Shabtai:2009:DMC:1550969.1551289}
A.~Shabtai, R.~Moskovitch, Y.~Elovici, C.~Glezer, Detection of malicious code
  by applying machine learning classifiers on static features: A
  state-of-the-art survey, Inf. Secur. Tech. Rep. 14~(1) (2009) 16--29.

\bibitem{SahuAhirwarHemlata2014}
M.~K. Sahu, M.~Ahirwar, A.~Hemlata, A review of malware detection based on
  pattern matching technique, Int. J. of Computer Science and Information
  Technologies (IJCSIT) 5~(1) (2014) 944--947.

\bibitem{Souri2018}
A.~Souri, R.~Hosseini, A state-of-the-art survey of malware detection
  approaches using data mining techniques, Human-centric Computing and
  Information Sciences 8~(1) (2018) 3.

\bibitem{LeDoux2015}
C.~LeDoux, A.~Lakhotia, Malware and machine learning, in: Intelligent Methods
  for Cyber Warfare, Springer, 2015, pp. 1--42.

\bibitem{Bazrafshan2013}
Z.~Bazrafshan, H.~Hashemi, S.~M.~H. Fard, A.~Hamzeh, A survey on heuristic
  malware detection techniques, in: Information and Knowledge Technology (IKT),
  2013 5th Conference on, IEEE, 2013, pp. 113--120.

\bibitem{Basu2016}
I.~Basu, Malware detection based on source data using data mining: A survey,
  American Journal Of Advanced Computing 3~(1).

\bibitem{Barriga2017MalwareDA}
J.~J. Barriga, S.~G. Yoo, Malware detection and evasion with machine learning
  techniques: A survey, International Journal of Applied Engineering Research
  12~(318).

\bibitem{Gardiner:2016:SML:2988524.3003816}
J.~Gardiner, S.~Nagaraja, On the security of machine learning in malware c\&c
  detection: A survey, ACM Comput. Surv. 49~(3) (2016) 59:1--59:39.

\bibitem{Shultz:2001}
M.~G. Schultz, E.~Eskin, F.~Zadok, S.~J. Stolfo, Data mining methods for
  detection of new malicious executables, in: Security and Privacy, 2001. S P
  2001. Proceedings. 2001 IEEE Symposium on, 2001, pp. 38--49.

\bibitem{Kolter:2006}
J.~Z. Kolter, M.~A. Maloof, Learning to detect and classify malicious
  executables in the wild, J. Mach. Learn. Res. 7 (2006) 2721--2744.

\bibitem{Ahmed2009}
F.~Ahmed, H.~Hameed, M.~Z. Shafiq, M.~Farooq, Using spatio-temporal information
  in api calls with machine learning algorithms for malware detection, in:
  Proceedings of the 2nd ACM workshop on Security and artificial intelligence,
  ACM, 2009, pp. 55--62.

\bibitem{Chau2010}
D.~H. Chau, C.~Nachenberg, J.~Wilhelm, A.~Wright, C.~Faloutsos, Polonium:
  Tera-scale graph mining for malware detection, in: ACM SIGKDD Conference on
  Knowledge Discovery and Data Mining, 2010, pp. 131--142.

\bibitem{FirdausiLimErwinEtAl2010}
I.~Firdausi, C.~Lim, A.~Erwin, A.~S. Nugroho, Analysis of machine learning
  techniques used in behavior-based malware detection, in: ACT '10, IEEE, 2010,
  pp. 201--203.

\bibitem{AndersonQuistNeilEtAl2011}
B.~Anderson, D.~Quist, J.~Neil, C.~Storlie, T.~Lane, Graph-based malware
  detection using dynamic analysis, Journal in Computer Virology 7~(4) (2011)
  247--258.

\bibitem{Santos:2011}
I.~Santos, J.~Nieves, P.~G. Bringas, International Symposium on Distributed
  Computing and Artificial Intelligence, Springer Berlin Heidelberg, Berlin,
  Heidelberg, 2011, Ch. Semi-supervised Learning for Unknown Malware Detection,
  pp. 415--422.

\bibitem{AndersonStorlieLane2012}
B.~Anderson, C.~Storlie, T.~Lane, Improving malware classification: bridging
  the static/dynamic gap, in: Proceedings of the 5th ACM workshop on Security
  and artificial intelligence, ACM, 2012, pp. 3--14.

\bibitem{Yonts2012}
J.~Yonts, Attributes of malicious files, Tech. rep., The SANS Institute (2012).

\bibitem{SantosDevesaBrezoEtAl2013}
I.~Santos, J.~Devesa, F.~Brezo, J.~Nieves, P.~G. Bringas, Opem: A
  static-dynamic approach for machine-learning-based malware detection, in:
  CISIS '12-ICEUTE{\'{}} 12-SOCO{\'{}}, Springer, 2013, pp. 271--280.

\bibitem{EskandariKhorshidpourHashemi2013}
M.~Eskandari, Z.~Khorshidpour, S.~Hashemi, Hdm-analyser: a hybrid analysis
  approach based on data mining techniques for malware detection, Journal of
  Computer Virology and Hacking Techniques 9~(2) (2013) 77--93.

\bibitem{Vadrevu2013}
P.~Vadrevu, B.~Rahbarinia, R.~Perdisci, K.~Li, M.~Antonakakis, Measuring and
  detecting malware downloads in live network traffic, in: Computer Security --
  ESORICS 2013: 18th European Symposium on Research in Computer Security,
  Egham, UK, September 9-13, 2013. Proceedings, Springer Berlin Heidelberg,
  Berlin, Heidelberg, 2013, pp. 556--573.

\bibitem{BaiWangZou2014}
J.~Bai, J.~Wang, G.~Zou, A malware detection scheme based on mining format
  information, The Scientific World Journal.

\bibitem{KruczkowskiSzynkiewicz2014}
M.~Kruczkowski, E.~N. Szynkiewicz, Support vector machine for malware analysis
  and classification, in: Web Intelligence (WI) and Intelligent Agent
  Technologies (IAT), IEEE Computer Society, 2014, pp. 415--420.

\bibitem{TamersoyRoundyChau2014}
A.~Tamersoy, K.~Roundy, D.~H. Chau, Guilt by association: large scale malware
  detection by mining file-relation graphs, in: Proceedings of the 20th ACM
  SIGKDD, ACM, 2014, pp. 1524--1533.

\bibitem{UppalSinhaMehraEtAl2014}
D.~Uppal, R.~Sinha, V.~Mehra, V.~Jain, Malware detection and classification
  based on extraction of api sequences, in: ICACCI, IEEE, 2014, pp. 2337--2342.

\bibitem{Chen:2015}
L.~Chen, T.~Li, M.~Abdulhayoglu, Y.~Ye, Intelligent malware detection based on
  file relation graphs, in: Semantic Computing (ICSC), 2015 IEEE International
  Conference on, 2015, pp. 85--92.

\bibitem{ElhadiMaarofBarry2015}
E.~Elhadi, M.~A. Maarof, B.~Barry, Improving the detection of malware behaviour
  using simplified data dependent api call graph, Journal of Security and Its
  Applications.

\bibitem{FengXiongCaoEtAl2015}
Z.~Feng, S.~Xiong, D.~Cao, X.~Deng, X.~Wang, Y.~Yang, X.~Zhou, Y.~Huang, G.~Wu,
  Hrs: A hybrid framework for malware detection, in: Proceedings of the 2015
  ACM International Workshop on Security and Privacy Analytics, ACM, 2015, pp.
  19--26.

\bibitem{Ghiasi:2015}
M.~Ghiasi, A.~Sami, Z.~Salehi, {Dynamic VSA: a framework for malware detection
  based on register contents }, Engineering Applications of Artificial
  Intelligence 44 (2015) 111 -- 122.

\bibitem{Ahmadi:2015}
M.~Ahmadi, G.~Giacinto, D.~Ulyanov, S.~Semenov, M.~Trofimov, Novel feature
  extraction, selection and fusion for effective malware family classification,
  CoRR abs/1511.04317.

\bibitem{Kwon2015}
B.~J. Kwon, J.~Mondal, J.~Jang, L.~Bilge, T.~Dumitras, The dropper effect:
  Insights into malware distribution with downloader graph analytics, in:
  Proceedings of the 22nd ACM SIGSAC Conference on Computer and Communications
  Security, ACM, 2015, pp. 1118--1129.

\bibitem{Mao2015}
W.~Mao, Z.~Cai, D.~Towsley, X.~Guan, Probabilistic inference on integrity for
  access behavior based malware detection, in: International Workshop on Recent
  Advances in Intrusion Detection, Springer, 2015, pp. 155--176.

\bibitem{Saxe2015}
J.~Saxe, K.~Berlin, Deep neural network based malware detection using two
  dimensional binary program features, in: Malicious and Unwanted Software
  (MALWARE), 2015 10th International Conference on, IEEE, 2015, pp. 11--20.

\bibitem{WuechnerOchoaPretschner2015}
T.~W{\"u}chner, M.~Ochoa, A.~Pretschner, Robust and effective malware detection
  through quantitative data flow graph metrics, in: Detection of Intrusions and
  Malware, and Vulnerability Assessment, Springer, 2015, pp. 98--118.

\bibitem{Raff2017}
E.~Raff, C.~Nicholas, An alternative to ncd for large sequences, lempel-ziv
  jaccard distance, in: Proceedings of the 23rd ACM SIGKDD International
  Conference on Knowledge Discovery and Data Mining, ACM, 2017, pp. 1007--1015.

\bibitem{Gharacheh:2015}
M.~Gharacheh, V.~Derhami, S.~Hashemi, S.~M.~H. Fard, Proposing an hmm-based
  approach to detect metamorphic malware, in: Fuzzy and Intelligent Systems
  (CFIS), 2015, pp. 1--5.

\bibitem{KhodamoradiFazlaliMardukhiEtAl2015}
P.~Khodamoradi, M.~Fazlali, F.~Mardukhi, M.~Nosrati, Heuristic metamorphic
  malware detection based on statistics of assembly instructions using
  classification algorithms, in: Computer Architecture and Digital Systems
  (CADS), 2015 18th CSI International Symposium on, IEEE, 2015, pp. 1--6.

\bibitem{UpchurchZhou2015}
J.~Upchurch, X.~Zhou, Variant: a malware similarity testing framework, in: 2015
  10th International Conference on Malicious and Unwanted Software (MALWARE),
  IEEE, 2015, pp. 31--39.

\bibitem{LiangPangDai2016}
G.~Liang, J.~Pang, C.~Dai, A behavior-based malware variant classification
  technique, International Journal of Information and Education Technology
  6~(4) (2016) 291.

\bibitem{Vadrevu2016}
P.~Vadrevu, R.~Perdisci, {MAXS: Scaling Malware Execution with Sequential
  Multi-Hypothesis Testing}, in: ASIA CCS '16, ASIA CCS '16, ACM, New York, NY,
  USA, 2016, pp. 771--782.

\bibitem{LeeMody2006}
T.~Lee, J.~J. Mody, Behavioral classification, in: EICAR Conference, 2006, pp.
  1--17.

\bibitem{HuangYeJiang2009}
K.~Huang, Y.~Ye, Q.~Jiang, Ismcs: an intelligent instruction sequence based
  malware categorization system, in: Anti-counterfeiting, Security, and
  Identification in Communication, 2009, IEEE, 2009, pp. 509--512.

\bibitem{ParkReevesMulukutlaEtAl2010}
Y.~Park, D.~Reeves, V.~Mulukutla, B.~Sundaravel, Fast malware classification by
  automated behavioral graph matching, in: Workshop on Cyber Security and
  Information Intelligence Research, ACM, 2010, p.~45.

\bibitem{Ye2010}
Y.~Ye, T.~Li, Y.~Chen, Q.~Jiang, Automatic malware categorization using cluster
  ensemble, in: Proceedings of the 16th ACM SIGKDD international conference on
  Knowledge discovery and data mining, ACM, 2010, pp. 95--104.

\bibitem{DahlStokesDengEtAl2013}
G.~E. Dahl, J.~W. Stokes, L.~Deng, D.~Yu, Large-scale malware classification
  using random projections and neural networks, in: Acoustics, Speech and
  Signal Processing (ICASSP), IEEE, 2013, pp. 3422--3426.

\bibitem{HuShinBhatkarEtAl2013}
X.~Hu, K.~G. Shin, S.~Bhatkar, K.~Griffin, Mutantx-s: Scalable malware
  clustering based on static features, in: USENIX Annual Technical Conference,
  2013, pp. 187--198.

\bibitem{IslamTianBattenEtAl2013}
R.~Islam, R.~Tian, L.~M. Batten, S.~Versteeg, Classification of malware based
  on integrated static and dynamic features, Journal of Network and Computer
  Applications 36~(2) (2013) 646--656.

\bibitem{Kong:2013}
D.~Kong, G.~Yan, Discriminant malware distance learning on structural
  information for automated malware classification, in: ACM SIGKDD '13, KDD
  '13, ACM, New York, NY, USA, 2013, pp. 1357--1365.

\bibitem{NariGhorbani2013}
S.~Nari, A.~A. Ghorbani, Automated malware classification based on network
  behavior, in: Computing, Networking and Communications (ICNC), 2013
  International Conference on, IEEE, 2013, pp. 642--647.

\bibitem{KawaguchiOmote2015}
N.~Kawaguchi, K.~Omote, Malware function classification using apis in initial
  behavior, in: Information Security (AsiaJCIS), 2015 10th Asia Joint
  Conference on, IEEE, 2015, pp. 138--144.

\bibitem{Lin:2015}
C.-T. Lin, N.-J. Wang, H.~Xiao, C.~Eckert, Feature selection and extraction for
  malware classification, Journal of Information Science and Engineering 31~(3)
  (2015) 965--992.

\bibitem{MohaisenAlrawiMohaisen2015}
A.~Mohaisen, O.~Alrawi, M.~Mohaisen, Amal: High-fidelity, behavior-based
  automated malware analysis and classification, computers \& security 52
  (2015) 251--266.

\bibitem{Pai:2015}
S.~Pai, F.~Di~Troia, C.~A. Visaggio, T.~H. Austin, M.~Stamp, Clustering for
  malware classification, J Comput Virol Hack Tech.

\bibitem{BaileyOberheideAndersenEtAl2007}
M.~Bailey, J.~Oberheide, J.~Andersen, Z.~M. Mao, F.~Jahanian, J.~Nazario,
  Automated classification and analysis of internet malware, in: Recent
  advances in intrusion detection, Springer, 2007, pp. 178--197.

\bibitem{BayerComparettiHlauschekEtAl2009}
U.~Bayer, P.~M. Comparetti, C.~Hlauschek, C.~Kruegel, E.~Kirda, Scalable,
  behavior-based malware clustering, in: NDSS, Vol.~9, 2009, pp. 8--11.

\bibitem{RieckTriniusWillemsEtAl2011}
K.~Rieck, P.~Trinius, C.~Willems, T.~Holz, Automatic analysis of malware
  behavior using machine learning, Journal of Computer Security 19~(4) (2011)
  639--668.

\bibitem{PalahanBabicChaudhuriEtAl2013}
S.~Palahan, D.~Babi{\'c}, S.~Chaudhuri, D.~Kifer, Extraction of statistically
  significant malware behaviors, in: Computer Security Applications Conference,
  ACM, 2013, pp. 69--78.

\bibitem{Egele:2014}
M.~Egele, M.~Woo, P.~Chapman, D.~Brumley, Blanket execution: Dynamic similarity
  testing for program binaries and components, in: USENIX Security '14, USENIX
  Association, San Diego, CA, 2014, pp. 303--317.

\bibitem{LindorferKolbitschComparetti2011}
M.~Lindorfer, C.~Kolbitsch, P.~M. Comparetti, Detecting environment-sensitive
  malware, in: Recent Advances in Intrusion Detection, Springer, 2011, pp.
  338--357.

\bibitem{SantosBrezoUgarte-PedreroEtAl2013}
I.~Santos, F.~Brezo, X.~Ugarte-Pedrero, P.~G. Bringas, Opcode sequences as
  representation of executables for data-mining-based unknown malware
  detection, Information Sciences 231 (2013) 64--82.

\bibitem{Polino:2015}
M.~Polino, A.~Scorti, F.~Maggi, S.~Zanero, Jackdaw: {Towards} {Automatic}
  {Reverse} {Engineering} of {Large} {Datasets} of {Binaries}, in: Detection of
  {Intrusions} and {Malware}, and {Vulnerability} {Assessment}, Lecture {Notes}
  in {Computer} {Science}, Springer International Publishing, 2015, pp.
  121--143.

\bibitem{Wong2006}
W.~Wong, M.~Stamp, Hunting for metamorphic engines, Journal in Computer
  Virology 2~(3) (2006) 211--229.

\bibitem{Attaluri2009}
S.~Attaluri, S.~McGhee, M.~Stamp, Profile hidden markov models and metamorphic
  virus detection, Journal in Computer Virology 5~(2) (2009) 151--169.

\bibitem{ChenRoussopoulosLiangEtAl2012}
Z.~Chen, M.~Roussopoulos, Z.~Liang, Y.~Zhang, Z.~Chen, A.~Delis, Malware
  characteristics and threats on the internet ecosystem, Journal of Systems and
  Software 85~(7) (2012) 1650--1672.

\bibitem{Comar:2013}
P.~M. Comar, L.~Liu, S.~Saha, P.~N. Tan, A.~Nucci, Combining supervised and
  unsupervised learning for zero-day malware detection, in: INFOCOM, 2013
  Proceedings IEEE, 2013, pp. 2022--2030.

\bibitem{Sexton2015}
J.~Sexton, C.~Storlie, B.~Anderson, {Subroutine based detection of APT
  malware}, Journal of Computer Virology and Hacking Techniques (2015) 1--9.

\bibitem{Park2009}
H.-S. Park, C.-H. Jun, A simple and fast algorithm for k-medoids clustering,
  Expert Systems with Applications 36 (2009) 3336 -- 3341.

\bibitem{Graziano:2015}
M.~Graziano, D.~Canali, L.~Bilge, A.~Lanzi, D.~Balzarotti, Needles in a
  haystack: Mining information from public dynamic analysis sandboxes for
  malware intelligence, in: USENIX Security '15, 2015, pp. 1057--1072.

\bibitem{Asquith2015}
M.~Asquith, Extremely scalable storage and clustering of malware metadata,
  Journal of Computer Virology and Hacking Techniques (2015) 1--10.

\bibitem{EST12}
M.~Egele, T.~Scholte, E.~Kirda, C.~Kruegel, A survey on automated dynamic
  malware-analysis techniques and tools, ACM computing surveys (CSUR) 44~(2)
  (2012) 6.

\bibitem{Siddiqui:2009}
M.~Siddiqui, M.~C. Wang, J.~Lee, Detecting internet worms using data mining
  techniques, Journal of Systemics, Cybernetics and Informatics (2009) 48--53.

\bibitem{Caliskan-Islam:2015}
A.~Caliskan-Islam, R.~Harang, A.~Liu, A.~Narayanan, C.~Voss, F.~Yamaguchi,
  R.~Greenstadt, De-anonymizing programmers via code stylometry, in: USENIX
  Security '15, USENIX Association, 2015, pp. 255--270.

\bibitem{SrakaewPiyanuntcharatsr:2015}
S.~Srakaew, W.~Piyanuntcharatsr, S.~Adulkasem, On the comparison of malware
  detection methods using data mining with two feature sets, Journal of
  Security and Its Applications 9 (2015) 293--318.

\bibitem{JangBrumleyVenkataraman2011}
J.~Jang, D.~Brumley, S.~Venkataraman, Bitshred: feature hashing malware for
  scalable triage and semantic analysis, in: Computer and communications
  security, ACM, 2011, pp. 309--320.

\bibitem{Allen:1970}
F.~E. Allen, Control flow analysis, in: Proceedings of a Symposium on Compiler
  Optimization, ACM, New York, NY, USA, 1970, pp. 1--19.

\bibitem{Kirat2013}
D.~Kirat, L.~Nataraj, G.~Vigna, B.~Manjunath, Sigmal: A static signal
  processing based malware triage, in: Proceedings of the 29th Annual Computer
  Security Applications Conference, ACM, 2013, pp. 89--98.

\bibitem{Tian:2008}
R.~Tian, L.~M. Batten, S.~C. Versteeg, Function length as a tool for malware
  classification, in: Malicious and Unwanted Software, 2008. MALWARE 2008. 3rd
  International Conference on, 2008, pp. 69--76.

\bibitem{Miramirkhani2017}
N.~Miramirkhani, M.~P. Appini, N.~Nikiforakis, M.~Polychronakis, Spotless
  sandboxes: Evading malware analysis systems using wear-and-tear artifacts,
  in: 2017 IEEE Symposium on Security and Privacy (SP), 2017, pp. 1009--1024.

\bibitem{Karpin2017}
J.~Karpin, A.~Dorfman, {Crypton - Exposing Malware's Deepest Secrets},
  \url{https://recon.cx/2017/montreal/resources/slides/RECON-MTL-2017-crypton.pdf},
  last accessed: 2018-10-18 (2017).

\bibitem{Blazytko2017}
T.~Blazytko, M.~Contag, C.~Aschermann, T.~Holz, Syntia: Synthesizing the
  semantics of obfuscated code, in: 26th {USENIX} Security Symposium ({USENIX}
  Security 17), {USENIX} Association, 2017, pp. 643--659.

\bibitem{Kotov2018}
V.~Kotov, M.~Wojnowicz, Towards generic deobfuscation of windows {API} calls,
  CoRR abs/1802.04466.

\bibitem{Liao2018}
Y.~Liao, R.~Cai, G.~Zhu, Y.~Yin, K.~Li, Mobilefindr: Function similarity
  identification for reversing mobile binaries, in: {ESORICS} {(1)}, Vol. 11098
  of Lecture Notes in Computer Science, Springer, 2018, pp. 66--83.

\bibitem{SurveySymExec-CSUR18}
R.~Baldoni, E.~Coppa, D.~C. D'Elia, C.~Demetrescu, I.~Finocchi, A survey of
  symbolic execution techniques, ACM Comput. Surv. 51~(3).

\bibitem{Guilfanov2008}
I.~Guilfanov, Decompilers and beyond, Black Hat USA.

\bibitem{Rosseau2018}
A.~Rosseau, R.~Seymour, {Finding Xori: Malware Analysis Triage with Automated
  Disassembly},
  \url{https://i.blackhat.com/us-18/Wed-August-8/us-18-Rousseau-Finding-Xori-Malware-Analysis-Triage-With-Automated-Disassembly.pdf},
  last accessed: 2018-10-14 (2018).

\bibitem{eschulte2018}
E.~Schulte, J.~Ruchti, M.~Noonan, D.~Ciarletta, A.~Loginov, Evolving exact
  decompilation, in: Binary Analysis Research (BAR), 2018, 2018.

\bibitem{vxheaven}
{Vxheaven}, \url{https://github.com/opsxcq/mirror-vxheaven.org}, accessed:
  2018-06-03.

\bibitem{offensive_computing}
{Offensive Computing}, \url{http://www.offensivecomputing.net}, accessed:
  2018-06-03.

\bibitem{malicia_project}
{Malicia Project}, \url{http://malicia-project.com}, accessed: 2018-06-03.

\bibitem{threattrack}
{ThreatTrack}, \url{https://www.threattrack.com/malware-analysis.aspx},
  accessed: 2018-06-03.

\bibitem{anubis}
{Anubis}, \url{https://seclab.cs.ucsb.edu/academic/projects/projects/anubis/},
  accessed: 2018-06-03.

\bibitem{BilgeBalzarottiRobertsonEtAl2012}
L.~Bilge, D.~Balzarotti, W.~Robertson, E.~Kirda, C.~Kruegel, Disclosure:
  detecting botnet command and control servers through large-scale netflow
  analysis, in: ACSAC '12, ACM, 2012, pp. 129--138.

\bibitem{Sanders2017}
H.~Sanders, { Garbage In, Garbage Out - How purportedly great ML models can be
  screwed up by bad data },
  \url{https://www.blackhat.com/docs/us-17/wednesday/us-17-Sanders-Garbage-In-Garbage-Out-How-Purportedly-Great-ML-Models-Can-Be-Screwed-Up-By-Bad-Data.pdf},
  last accessed: 2018-10-18 (2017).

\bibitem{Miller2015}
B.~Miller, A.~Kantchelian, S.~Afroz, R.~Bachwani, R.~Faizullabhoy, L.~Huang,
  V.~Shankar, M.~Tschantz, T.~Wu, G.~Yiu, et~al., Back to the future: Malware
  detection with temporally consistent labels, CORR.

\bibitem{Jordaney2017}
R.~Jordaney, K.~Sharad, S.~K. Dash, Z.~Wang, D.~Papini, I.~Nouretdinov,
  L.~Cavallaro,
  \href{https://www.usenix.org/conference/usenixsecurity17/technical-sessions/presentation/jordaney}{Transcend:
  Detecting concept drift in malware classification models}, in: 26th {USENIX}
  Security Symposium ({USENIX} Security 17), {USENIX} Association, Vancouver,
  BC, 2017, pp. 625--642.
\newline\urlprefix\url{https://www.usenix.org/conference/usenixsecurity17/technical-sessions/presentation/jordaney}

\bibitem{Harang2018}
R.~Harang, F.~Ducau, {Measuring the Speed of the Red Queens Race},
  \url{https://i.blackhat.com/us-18/Wed-August-8/us-18-Harang-Measuring-the-Speed-of-the-Red-Queens-Race.pdf},
  last accessed: 2018-10-18 (2018).

\bibitem{virtustotal}
{VirusTotal}, \url{https://www.virustotal.com}, accessed: 2018-06-03.

\bibitem{malwr}
{Malwr}, \url{https://malwr.com}, accessed: 2018-06-03.

\bibitem{Ruthven2017}
M.~Ruthven, A.~Blaich, {Fighting targeted malware in the mobile ecosystem},
  \url{https://www.blackhat.com/docs/us-17/wednesday/us-17-Ruthven-Fighting-Targeted-Malware-In-The-Mobile-Ecosystem.pdf},
  last accessed: 2018-10-16 (2017).

\bibitem{10.1007/978-3-319-60080-2_21}
G.~Laurenza, L.~Aniello, R.~Lazzeretti, R.~Baldoni, Malware triage based on
  static features and public apt reports, in: S.~Dolev, S.~Lodha (Eds.), Cyber
  Security Cryptography and Machine Learning, Springer International
  Publishing, Cham, 2017, pp. 288--305.

\bibitem{8323965}
M.~Howard, A.~Pfeffer, M.~Dalai, M.~Reposa, Predicting signatures of future
  malware variants, in: 2017 12th International Conference on Malicious and
  Unwanted Software (MALWARE), 2017, pp. 126--132.

\bibitem{juzonis2012specialized}
V.~Juzonis, N.~Goranin, A.~Cenys, D.~Olifer, Specialized genetic algorithm
  based simulation tool designed for malware evolution forecasting, Annales
  Universitatis Mariae Curie-Sklodowska, sectio AI--Informatica 12~(4).

\bibitem{Hal:2012}
H.~Pomeranz, Detecting malware with memory forensics,
  \url{http://www.deer-run.com/~hal/Detect_Malware_w_Memory_Forensics.pdf},
  last accessed: 2018-05-14 (2012).

\bibitem{damodaran2015comparison}
A.~Damodaran, F.~Di~Troia, C.~A. Visaggio, T.~H. Austin, M.~Stamp, A comparison
  of static, dynamic, and hybrid analysis for malware detection, Journal of
  Computer Virology and Hacking Techniques (2015) 1--12.

\bibitem{Kulakov2017}
Y.~Kulakov, Mazewalker,
  \url{https://recon.cx/2017/montreal/resources/slides/RECON-MTL-2017-MazeWalker.pdf}
  (2017).

\bibitem{Laurenza2016}
G.~Laurenza, D.~Ucci, L.~Aniello, R.~Baldoni, An architecture for
  semi-automatic collaborative malware analysis for cis, in: Dependable Systems
  and Networks Workshop, 2016 46th Annual IEEE/IFIP International Conference
  on, IEEE, 2016, pp. 137--142.

\end{thebibliography}

\end{document}